\documentclass[%
 aip,
superscriptaddress,
preprintnumbers,
nofootinbib,
showkeys,
floatfix,
eqsecnum,
amsmath,amssymb,
 reprint,%
]{revtex4-1}
\usepackage[utf8]{inputenc}
\usepackage[T1]{fontenc}
\usepackage{mathptmx}
\DeclareMathAlphabet{\mathcal}{OMS}{cmsy}{m}{n}
\usepackage{bm}
\usepackage{booktabs}
\usepackage{placeins}
\usepackage{multirow}
\usepackage{hyphenat}
\usepackage{graphicx}%

\usepackage{caption}
\usepackage[justification=centerlast]{subcaption}

\usepackage{dcolumn}
\newcolumntype{V}{!{\vrule width 1.5pt}} 
\usepackage{makecell}
\usepackage{etoolbox}

\newcolumntype{C}[1]{>{\centering\arraybackslash}p{#1}}

\usepackage{algorithm}
\usepackage{algorithmicx}
\usepackage{algpseudocode}
\algrenewcommand\algorithmicrequire{\textbf{Input:}}
\algrenewcommand\algorithmicensure{\textbf{Output:}}
\usepackage{afterpage}

\usepackage[colorlinks=true,
            linkcolor=blue,
            urlcolor=blue,
            citecolor=blue]{hyperref}
\usepackage{cleveref}
\crefname{figure}{Fig.}{Fig.}
\Crefname{figure}{Fig.}{Fig.}
\crefname{section}{Section}{Sections}
\crefname{table}{Table}{Tables}
\crefname{equation}{Eq.}{Eqs.}

\makeatletter
\def\@email#1#2{%
 \endgroup
 \patchcmd{\titleblock@produce}
  {\frontmatter@RRAPformat}
  {\frontmatter@RRAPformat{\produce@RRAP{*#1\href{mailto:#2}{#2}}}\frontmatter@RRAPformat}
  {}{}
}%
\makeatother

\usepackage{soul}
\usepackage[usenames,dvipsnames]{xcolor}
\usepackage{xspace}

\newcommand{\sigmoid}{\tau}
\newcommand{\ucb}{\text{ucb}}

\DeclareMathOperator*{\argmax}{arg\max}
\newcommand{\bx}{\bm{x}}
\newcommand{\runit}{\Omega/\text{sq}}

\begin{document}

\date{\today}


\title{Bayesian Optimization for Stable Properties Amid Processing Fluctuations in Sputter Deposition}

\author{Ankit Shrivastava}
    \email{ashriva@sandia.gov}
    \affiliation{Sandia National Laboratories}
\author{Matias Kalaswad}
    \email{mskalas@sandia.gov}
    \affiliation{Sandia National Laboratories}
\author{Joyce O. Custer}
    \email{jcuster@sandia.gov}
    \affiliation{Sandia National Laboratories}
\author{David P. Adams}
    \email{dpadams@sandia.gov}
    \affiliation{Sandia National Laboratories}
\author{Habib N. Najm}
    \email{hnnajm@sandia.gov}
    \affiliation{Sandia National Laboratories}

\begin{abstract}
We introduce a Bayesian optimization approach to guide the sputter deposition of molybdenum thin films, aiming to achieve desired residual stress and sheet resistance while minimizing susceptibility to stochastic fluctuations during deposition. Thin films are pivotal in numerous technologies, including semiconductors and optical devices, where their properties are critical. Sputter deposition parameters, such as deposition power, vacuum chamber pressure, and working distance, influence physical properties like residual stress and resistance. Excessive stress and high resistance can impair device performance, necessitating the selection of optimal process parameters. Furthermore, these parameters should ensure the consistency and reliability of thin film properties, assisting in the reproducibility of the devices. However, exploring the multidimensional design space for process optimization is expensive. Bayesian optimization is ideal for optimizing inputs/parameters of general black-box functions without reliance on gradient information. We utilize Bayesian optimization to optimize deposition power and pressure using a custom-built objective function incorporating observed stress and resistance data. Additionally, we integrate prior knowledge of stress variation with pressure into the objective function to prioritize films least affected by stochastic variations. Our findings demonstrate that Bayesian optimization effectively explores the design space and identifies optimal parameter combinations meeting desired stress and resistance specifications.

\end{abstract}
\maketitle
\section{Introduction}\label{sec:intro}

Recent advances in computational, machine learning and statistical methods have significantly impacted the capabilities for optimally designing metallurgical thin films.
Thin films are used in various microelectronics, optics, and mechanical applications ranging from integrated circuits to reflective or protective coatings. 
Among metallurgical thin films, refractory metals are beneficial due to their remarkable resistance to heat and wear. 
Their high melt temperatures, low diffusivity, and good electrical conductivity often make them ideal for applications involving elevated temperatures and electronics. 
Thermally stable contacts, Ohmic contacts, and interconnects are prevalent in the semiconductor industry, and all capitalize on such properties of refractory metal thin films and coatings \cite{abramsky2012other,messica1995refractory,erofeev2018high}.
Notably, refractory thin films can be fabricated using various methods, such as chemical vapor deposition, sputter deposition, and vacuum thermal evaporation.
The requisite process steps, conditions, and tolerances play a crucial role in achieving target properties, such as low resistance during the fabrication of thin films for specific applications.

Determining the best process conditions to fabricate thin films with desired properties requires optimizing the process-property models. 
Often, information about such models is limited, necessitating exploring the process conditions space.
Exploration based on expert intuition and grid search \cite{yasrebi2014optimization} has been used for optimizing process conditions for thin films.
However, these methods may require many experiments to be carried out, depending on the dimension of the process and property parameters (design space), making them time-consuming and costly. 
Due to experimental budget constraints, an efficient design of experiments (DoE) to search for the optimum is imperative.
In addition to producing thin films with target properties, the optimal properties achieved must remain robust to unavoidable process fluctuations. 
This robustness is essential to ensure consistently good process target properties, sometimes at the expense of some sub-optimality. 
A highly optimized but non-robust process output is not a desirable setup for a manufacturing context.

Over the years, various DoE methods \cite{yucel2014optimization, dreer1999multidimensional, ramana2012optimization} have been proposed to guide toward optimum process conditions sequentially.
These DoE methods involve iteratively exploring sampled processing conditions and learning a process-property surrogate model (response surface) to determine the next set of process conditions for experiments. 
DoE methods typically sample process conditions using Latin hypercube sampling, full factorial \cite{fisher1936design, madhava1956sequential}, or central composite\cite{box1951experimental} to learn a response surface.
For robust design, methods such as the Taguchi method\cite{taguchi1989quality} have been proposed to handle experimental noise.
The extension of DoE methods for constraint optimizations has also been suggested in work such as that by Jones et al. \cite{jones1998efficient}.
However, these methods learn a predefined parametric form, such as linear and quadratic surfaces, which limits their applicability to complex systems.
Additionally, these methods tend to be exploitative and often converge to a local optimum, requiring extensive parameter space exploration to reach the global optimum.
Extensive exploration increases the number of trials exponentially, making classical DoE methods costly for resource-intensive experiments such as sputtering deposition.

Data-driven Bayesian optimization (BayesOpt) algorithms have been found to model complex systems and explore the process conditions more efficiently\cite{packwood2017bayesian, garnett2023bayesian}.
Unlike classical DoE methods, BayesOpt employs a probability distribution over functions using nonparametric models such as the Gaussian process.
The flexibility of this nonparametric model enables learning complex nonlinear relationships.
Furthermore, the probability distribution quantifies uncertainties, which help differentiate the regions of parameter space as explored and unexplored, helping to balance exploitation and exploration.
Balancing exploitation and exploration helps prevent locally optimal solutions and suboptimal searches, leading to faster convergence to a global optimum.
Due to its probabilistic structure, BayesOpt can also handle experimental noise with fewer trials than the Taguchi method. 
Further, various BayesOpt algorithms\cite{daulton2022robust,
sanders2019bayesian, le2020bayesian, nguyen2021value} have been proposed to obtain robust process output efficiently.

In materials manufacturing, Bayesian optimization has guided experiments and optimized process conditions for desired properties. 
It has been demonstrated that BayesOpt converges faster than traditional DoE methods in multiple material science applications~\cite{liang2021benchmarking, lookman2019active}.
In particular, BayesOpt has optimized thin film growth targeting desired properties. 
Studies have successfully optimized crystal structures~\cite{ohkubo2021realization}, optical contrast~\cite{kusne2020fly}, residual resistivity ratio~\cite{wakabayashi2019machine}, and contact resistance~\cite{miyagawa2021application}. 
In recent years, various strategies based on BayesOpt have been suggested to address challenging optimization problems in material science, such as multi-objective problems~\cite{gopakumar2018multi,khatamsaz2022multi}, and mixed-variable problems with robust design~\cite{zhang2020bayesian}.

In this work, we present a Bayesian optimization construction to identify process conditions that satisfy multiple property constraints for thin films while simultaneously maximizing optimized process robustness against minor variations in sputter deposition conditions.
To our knowledge, existing work for optimizing thin film deposition has not addressed the challenge of concurrently identifying process conditions that aim to satisfy both optimality and robustness.

The proposed approach is tested on molybdenum (Mo) thin films, a well-studied refractory metal, with results readily validated against existing literature.
Coatings with Mo have practical uses, such as in Mo/Si multilayers for extreme-UV optics \cite{barbee1985molybdenum, mirkarimi1999stress} and back contacts in CIGS solar cells \cite{orgassa2003alternative, li2016analysis}. 
In the latter application, and often generally, the Mo layer should have low electrical resistance and residual stress to ensure good adhesion to other layers. 
We used these criteria to build an objective function for the Bayesian optimization of Mo films presented here.

In our BayesOpt guided experiment, we fabricated Mo thin films using sputter deposition, a well-established technique compatible with refractory metal microelectronics. 
We deposited a range of films with various sputter powers and argon (Ar) pressures – two process parameters that typically significantly influence film properties. 
We then used data from measurements of these fundamental properties (e.g., resistance, stress) to update the process-property model and calculate the next set of deposition conditions to explore. 
We repeated this process until a stable set of conditions was found that produced Mo thin films with the desired properties. 

We observed that BayesOpt can guide the sputter deposition experiment efficiently. 
With a few additional experiments and prior knowledge, the algorithm explored feasible power and pressure values, ultimately discovering the optimal configuration that satisfied the desired criteria for Mo films. 
It was also observed that along with being robust to experimental noise during deposition, our BayesOpt construction was also robust to unquantified environmental uncertainties due to deposition chamber cleaning.

The remainder of this article is organized as follows.
In section \ref{sec:criteria}, we discuss the design objectives for Mo thin film. 
In section \ref{sec:method}, we elucidate the Bayesian optimization methodology employed to achieve the design specifications. 
Section \ref{sec:Ex_setup} delves into the details of our experimental setup. 
Section \ref{sec:results} presents the obtained results, and in section \ref{sec:conclusion}, we draw our conclusions from the study. 

\section{Design specifications for Thin Films} \label{sec:criteria}
We aim to discover power and pressure conditions for sputter deposition of Mo thin films such that they meet the following design specifications (criteria):
\begin{enumerate}
    \item The first criterion requires that the films exhibit residual stresses in the range of ($-300$ MPa, $300$ MPa). 
    For Mo thin films, this range is considered acceptable for low residual stress.
    A low residual stress criterion is crucial for ensuring the integrity of the resulting thin film against various failure mechanisms. 
    Like almost all thin film applications, films must adhere to a substrate or underlying surface. 
    Films with large residual stresses are more susceptible to failure mechanisms such as buckling, cracking, and delamination – phenomena detrimental to the device or coating's overall function{\cite{abadias2018stress}}.
    
    \item The second criterion specifies that the films exhibit resistance below $3$ $\runit$, which falls within an acceptable range for low sheet resistance.
    In most microelectronic applications, it is also crucial that metal layers have low resistance to minimize losses and resistive heating.
    
    \item The third criterion requires that the films be as dense as possible in a qualitative sense. 
    Density plays a crucial role in critical applications, such as microelectronics, where low-density metal layers lead to higher resistance (pores/voids restrict electrical paths).
    We enforce this criterion by favoring design configurations where the derivative of stress with Ar pressure is positive.
    This criterion is supported by the behavior of magnetron-sputtered metal films.
    These films demonstrate a compressive-to-tensile stress transition with increasing process pressure. 
    Films with stress near this transition (characterized by a large, positive slope) are generally denser than those deposited at higher pressures, assuming other factors are held similar. 
    At Ar pressures greater than this transition point --- where the derivative of stress becomes negative --- films are typically underdense and porous{\cite{abadias2018stress}}.
    
    \item Finally, the fourth objective requires that the films' stress be least susceptible to pressure fluctuations.
    As already discussed, sputtered metal films can undergo a large change in stress over a narrow range of process conditions. 
    Film stress can be particularly sensitive to sputter pressure. 
    We target a robust set of process conditions where a slightly different pressure causes a minimal change in the film stress. 
    We achieve this objective by minimizing the derivative of the stress with Ar pressure.
    This objective helps address the manufacturing challenge of finding deposition conditions where the properties of a sputter-deposited metal are robust to process fluctuations.  
\end{enumerate}

\section{Optimization for Design Specifications}\label{sec:method}
We use BayesOpt to iteratively search for the optimal power and pressure pairs within the bounded domain of feasible sputter deposition configurations, denoted by $\mathcal{X} \subset \mathbb{R}_{\geq 0}^2$ such that they satisfy the design specifications as discussed in \cref{sec:criteria}. 

In the following, we introduce our notation. During each iteration of experiments, thin films are deposited using a given process condition (design point) $\bx = (x_\text{po}, x_\text{pr}) \in \mathcal{X}$, where the subscript ``po" and ``pr" denotes the power and pressure dimension respectively. 
Subsequently, their stress $S(\bx)$ and sheet resistance  $R(\bx)$ are measured.
The set of $n$ power and pressure pairs at which thin film measurements have been collected is denoted as $\mathcal{D} = [\bx_1, \bx_2, \dots, \bx_n]$, where $\mathcal{D}\subset\mathcal{X}$. 
The set of the outputs of any function $y(\bx)$ at the observed pairs are denoted as $y(\mathcal{D}) = [y(\bx_{1}), y(\bx_{2}) \dots y(\bx_{n})]^T$.
The set of collected property measurements with corresponding process conditions $\{\mathcal{D}, S(\mathcal{D}), R(\mathcal{D})\}$ is referred to as the observed dataset.

The primary challenge lies in the fact that the stress $S(\bx)$ and sheet resistance $R(\bx)$ functions are unknown (black box). 
The only information available is the set of measurements of stress, $S(\mathcal{D})$ and sheet resistance $R(\mathcal{D})$ at observed design points $\mathcal{D}$.
Using the available information, we first obtain an objective function  (\cref{sec:objective_function}), which we then optimize using the Bayesian optimization algorithm (\cref{sec:bayes_opt}) to satisfy the design specifications.
The method is summarized in algorithm \ref{alg:bayesian_optimization}.

\afterpage{
\begin{algorithm}[H]
\begin{algorithmic}[1]
\Require{Set of prior sputter configurations $\mathcal{D}$, stress and sheet resistance values $\mathcal{M} = (S(\mathcal{D}), R(\mathcal{D}))$. Predefined $\epsilon_{\text{po}}$, $\epsilon_{\text{pr}}$  and stagnation limit $N$.}
\Ensure{Optimal solution $\bx^*$ and maximum value $f(\bx^*)$}
\State Initialize stagnation count, $n = 0$
\While{$n < N$}
  \State Compute $f_1(\mathcal{D}) \dots f_4(\mathcal{D})$ using $\mathcal{M}$ in \cref{eq:stress_criteria}-\cref{eq:min_grad_criteria}.
  \State Compute $f(\mathcal{D})$, \cref{eq:combined}.
  \State Compute $\mu_{\text{pos}}(\bx)$ and $\sigma_{\text{pos}}(\bx)$, \cref{eq:posterior}. 
  \State Compute the optimal solution $\bx^*$ using \cref{eq:optimum}.
  \vspace{0.25em}
  \If{$|x^*_{\text{old}, \text{pr}} - x_{\text{pr}}^*| < \epsilon_{\text{pr}}$ and $|x^*_{\text{old}, \text{po}} - x_{\text{po}}^*| < \epsilon_{\text{po}}$}
        \State $n++$
    \Else
        \State $n=0$
    \EndIf
  \vspace{0.25em}
  \State Store : $\bx^*_{\text{old}} \leftarrow \bx^*$
  \State Compute the next design point $\bar{\bx}$ using \cref{eq:next_point}.
  \State Deposit thin films with process conditions $\bar{\bx}$.
  \State Measure stress $\bar{s} = S(\bar{\bx})$ and sheet resistance $\bar{r} = R(\bar{\bx})$
  \State Update the current dataset: $$\mathcal{D} \leftarrow \mathcal{D} \cup \{\bar{\bx}\} \quad \text{and} \quad \mathcal{M} \leftarrow \mathcal{M} \cup \{(\bar{s}, \bar{r})\}$$
\EndWhile
\State Report $\bx^*$ as optimal solution.
\end{algorithmic}
\caption{Bayesian Optimization of Sputter Deposition Configuration.}\label{alg:bayesian_optimization}
\end{algorithm}
}

\subsection{Obtaining a Unified Design Objective} \label{sec:objective_function}

The choice of the objective function is made such that it inherently accounts for all design specifications.
First, we start by selecting four distinct smooth functions, $f_1,f_2,f_3,f_4:\mathcal{X} \rightarrow \mathbb{R}$, that satisfy the four design specifications individually. 
Afterward, we combine the separate functions into a single unified smooth function, $f$, that serves as an objective for our optimization.
Selecting smooth functions helps effectively fit the objective function by the surrogate model in the Bayesian algorithm. The four functions are defined in the following.

With $T_s$ defined as a threshold stress magnitude, we employ the first objective $f_1(\bx)$ to enforce the requirement that the stress satisfies $-T_s \leq S(\bx) \leq T_s$. Thus, $f_1(\bx)$ is defined as
\begin{align}
f_1(\bx) &= \sigmoid (T_s + S(\bx)) + \sigmoid (T_s - S(\bx)) \label{eq:stress_criteria}
\end{align}
where
\begin{align}
\sigmoid(z) &= \frac{1}{1+\exp^{-z}} \quad z \in \mathbb{R} \notag
\end{align}
is the standard logistic function, which maps its input to a range between $0$ and $1$ with a smooth step-like transition. This objective maps the stress values to $(0,1)$ such that $f_1(\bx)\simeq 1$ for $S(\bx) \in [-T_s, T_s]$ and $0$ elsewhere.

The second objective requires the films to have $R(\bx) < T_r$, where $T_r$ is a specified threshold sheet resistance value. This objective is enforced using $f_2(\bx)$, a modified form of the logistic function, given as
\begin{align}
f_2(\bx) &= \frac{1}{1+\exp^{-m(T_r-R(\bx))}}, \label{eq:resistance_criteria}
\end{align}
where the scalar $m$ controls the transition slope in range $(0, 1)$. 
This function maps $R(\bx)$ values below $T_r$ to $1$, and $0$ elsewhere.

The third objective, requiring the thin films to be dense, is enforced by ensuring that the derivative of stress with pressure is positive, $\frac{\partial S(\bx)}{\partial x_\text{pr}} > 0$, as discussed in Section {\ref{sec:criteria}}.
Since the function of stress $S(\bx)$ is unknown, to obtain the derivative, we first fit a Gaussian process (GP) mean interpolant $g(\bx)$ over stress observations and then compute its derivative with respect to pressure.  
Selecting the GP mean as an interpolant eliminates the need to make assumptions about stress's functional form. The details about the GP construction are provided in \cref{sec:gaussian_process}.
The GP mean function is given by
\begin{align}
 S(\bx) \approx g(\bx)  &= k_{\phi}(\bx, \mathcal{D}) (k_{\phi}(\mathcal{D}, \mathcal{D})+\sigma^2_\text{noise}I_n)^{-1} S(\mathcal{D}), \label{eq:interpolant}
\end{align}
where $I_n=\text{diag}(1,\ldots,1)\in\mathbb{R}^{n\times n}$, we presume a zero GP prior mean function, additive Gaussian independent identically distributed data noise with zero mean and $\sigma^2_{\text{noise}}$ variance, and $k_\phi(\cdot,\cdot)$ is a squared exponential radial basis function (RBF) kernel with hyperparameters $\phi$. The kernel structure follows the formulation defined in Eq.~\ref{eq:kernel}, \cref{sec:hyper_tune}, and its hyperparameters are tuned similarly. 
It is important to note that $g(\bx)$ differs from the surrogate model utilized in the context of Bayesian optimization as described in \cref{sec:bayes_opt}. 
To ensure clarity, we will use the term 'interpolant' to represent $g(\bx)$ for the rest of the article.
With this surrogate for $S(\bx)$ in hand, the 3$^{rd}$ objective is met using a hyperbolic tangent function, and $f_3(\bx)$, defined as 
\begin{align}
f_3(\bx) &= \frac{\tanh(g_{\text{pr}}(\bx)) +1}{2} \label{eq:pos_grad_criteria} ;\quad g_{\text{pr}}(\bx) := \frac{\partial g(\bx)}{\partial x_{\text{pr}}}.
\end{align}
where, $f_3(\bx)$ takes the derivative of stress with pressure as input and maps to $1$ for the nonnegative input and $0$ elsewhere.

The final objective requires minimal film stress dependence on pressure fluctuation.
To achieve this, we maximize the function $f_4(\bx)$, defined as 
\begin{align}
f_4(\bx) &= \frac{-g_{\text{pr}}(\bx)}{B} + 1 , \label{eq:min_grad_criteria}
\end{align}
where $B$ is a constant scale factor. This function output increases linearly with decreasing $g_{\text{pr}}(\bx)$. We enforce the $f_4(\bx)$ range to be roughly between $0$ and $1$, aligning it in scale with the other three functions, with a suitable choice of $B$.

Finally, we define the composite function $f(\bx)$ as
\begin{align}
f(\bx) &= f_1(\bx)f_2(\bx)f_3(\bx)f_4(\bx).\label{eq:combined}
\end{align}
The overall output is influenced by all functions, effectively taking care of all four criteria. The composite function $f(\bx)$ is the objective function for our optimization problem. 

\begin{figure*}[htbp]
    \centering
    \begin{subfigure}{17.9pc}    
        \includegraphics[width=\textwidth]{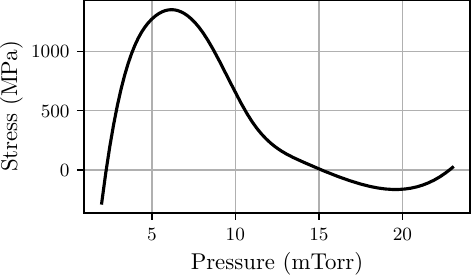}
        \caption{stress, $S(\bx)$  vs pressure}
        \label{fig:stress_example}
    \end{subfigure}
    \hspace{1pc}
    \begin{subfigure}{17.9pc}
        \includegraphics[width=\textwidth]{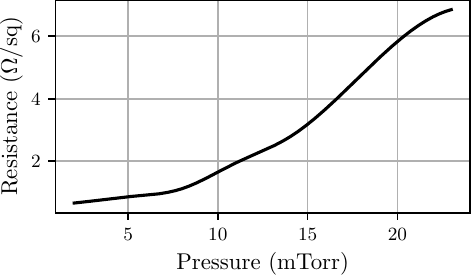}
        \caption{resistance, $R(\bx)$ vs pressure}
        \label{fig:resistance_example}
    \end{subfigure}

    \begin{subfigure}{17.9pc}
        \vspace{1em}
        \includegraphics[width=\textwidth]{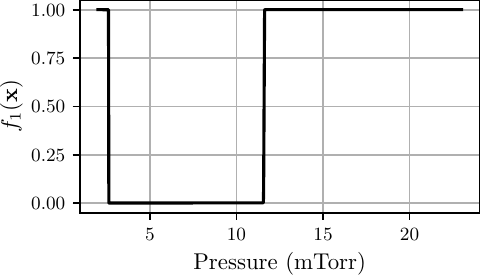}
        \caption{stress criterion, $f_1(\bx)$ vs pressure}
        \label{fig:criteria_1}
    \end{subfigure}
    \hspace{1pc}
    \begin{subfigure}{17.9pc}
        \vspace{1em}
        \includegraphics[width=\textwidth]{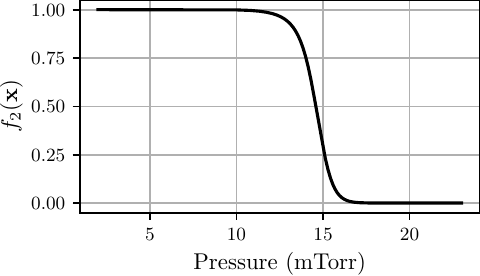}
        \caption{resistance criterion, $f_2(\bx)$ vs pressure}
        \label{fig:criteria_2}
    \end{subfigure}

    \begin{subfigure}{17.9pc}
        \vspace{1em}
        \includegraphics[width=\textwidth]{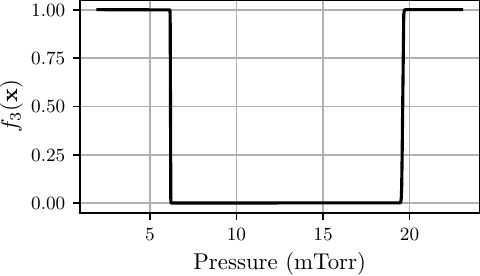}
        \caption{postive derivative criterion, $f_3(\bx)$ vs pressure}
        \label{fig:criteria_3}
    \end{subfigure}
    \hspace{1pc}
    \begin{subfigure}{17.9pc}
        \vspace{1em}
        \includegraphics[width=\textwidth]{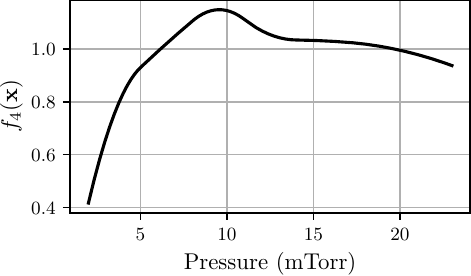}
        \caption{minimum derivative criterion, $f_4(\bx)$ vs pressure}
        \label{fig:criteria_4}
    \end{subfigure}
    
    \begin{subfigure}{17.9pc}
    \vspace{1em}
    \includegraphics[width=\textwidth]{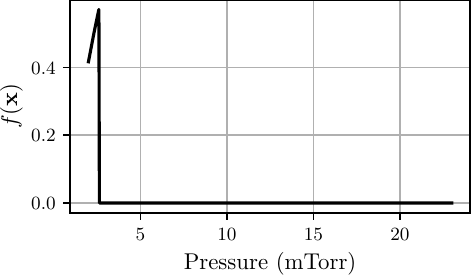}
    \caption{unified objective function, $f(\bx)$ vs pressure}
    \label{fig:combined}
    \end{subfigure}
    \caption{
   Figures \ref{fig:stress_example} and \ref{fig:resistance_example} illustrate the stress and resistance variations with pressure for a fixed power, respectively.
   For the given stress and resistance profiles, the behavior of the four design criteria functions $f_1$, $f_2$, $f_3$, and $f_4$ are shown in Figs.~{\ref{fig:criteria_1}}, {\ref{fig:criteria_2}}, {\ref{fig:criteria_3}}, {\ref{fig:criteria_4}}, respectively. 
   The pressure values at which these functions reach their maximum satisfy the corresponding design criteria. 
   The pressure that maximizes the unified objective function $f$, shown in \cref{fig:combined}, satisfies all design criteria.
   }
\label{fig:criteria_illustration}
\end{figure*}

We have illustrated the behavior of these functions in \cref{fig:criteria_illustration}, where we offer an example of stress and sheet resistance in a one-dimensional pressure-variation context for a fixed power condition. 
The stress and sheet resistance dependence on pressure is shown in Figs. \ref{fig:stress_example} and \ref{fig:resistance_example}.
The corresponding profiles of $f_1, f_2, f_3, f_4$ are shown in Figs.~\ref{fig:criteria_1}, \ref{fig:criteria_2}, \ref{fig:criteria_3}, and \ref{fig:criteria_4}, respectively. 
The final objective function $f$ is shown in \cref{fig:combined}. 
Notably, for this example, the maximum of $f$ lies at a pressure value of $2$ mTorr for the given power setting, indicating the deposition conditions that satisfy the design criteria. 

Note that the four design criteria can be conceptualized as elements of a constrained optimization problem, with the first three criteria formulated as constraints and the fourth used as the objective function.
One way to solve this problem involves using a Lagrangian approach, where constraints are included with penalties{\cite{gramacy2016modeling}} within the objective function, transforming the problem into an unconstrained optimization task.
This adjustment removes the strict criteria of the original constraints, opting instead to penalize deviations from them, effectively relaxing the constraint criteria.
Several extensions of BayesOpt algorithms have also been proposed to solve constrained optimization problems where the methods leverage variants of acquisition functions such as expected improvement\cite{gardner2014bayesian, 
feliot2017bayesian} and predictive entropy search with constraints\cite{hern2016general}.
However, these variants of acquisition functions for constrained problems can be costly for resource-intensive experiments, as they can result in the rejection of exploitation-based proposals.
Given the limited experimental budget, we opted to formulate the problem in a manner inspired by filter-based optimization approaches\cite{pourmohamad2020statistical, sainvitu2007filter}. 
The proposed approach prioritizes the search toward the feasible space, reducing the number of trials required to achieve the optimum.

\subsection{Bayesian Optimization Algorithm} \label{sec:bayes_opt}
The BayesOpt algorithm aims to find a design that optimizes the objective function $f(\cdot)$ by iteratively evaluating $f(\cdot)$ at existing design points and using these evaluations, $f(\mathcal{D})$ to propose new design points.
The process involves two tasks: (1) building a probabilistic surrogate model from the existing observations; see \cref{sec:gaussian_process}, and (2) using an acquisition function that guides experiments where to sample next in search space; see \cref{sec:ucb}. The iteration process continues until a termination criterion is achieved; see \cref{sec:terminate}. 

\subsubsection{Building Surrogate Model of Objective Function} \label{sec:gaussian_process}
We use a Gaussian process~\cite{Rasmussen:2006} (GP) as the surrogate model, which, when fitted to $(\mathcal{D},f(\mathcal{D}))$, provides a probabilistic prediction of $f(\bx)$ at all points $\bx\in\mathcal{X}$. 
This information is used in the acquisition function to sample the next point. 
GPs are nonparametric models, which makes them highly flexible to fit complex mappings. 

A GP is a probability distribution over random functions that are jointly Gaussian at any finite set of points. It is defined by its mean function $\mu(\bx)$ and covariance function $k(\bx,\bx')$. 
In our present setting, we define the covariance function as a squared exponential kernel $k_\theta(\bx,\bx')$, specified later in this section so that we write the GP as
\begin{align}
	\hat{f}(\bx) &\sim \mathcal{GP}(\mu(\bx), k_{\theta}(\bx, \bx')) \quad \forall \bx \in \mathcal{X}.
\end{align}  
Here, $\hat{f}$ is a surrogate function of $f$, sampled from the Gaussian process $\mathcal{GP}$. 
Using observed data independent coordinates $\mathcal{D}=[\bx_1,\ldots,\bx_n]$, we define a prior distribution 
\begin{equation}    
	\mathcal{GP}(\mu(\mathcal{D}), k_{\theta}(\mathcal{D}, \mathcal{D})), \label{eq:prior}
\end{equation}
where $\mu(\mathcal{D}) = [\mu(\bx_{1}), \mu(\bx_{2}) \dots \mu(\bx_{n})]$ is the prior mean vector and $k_\theta(\mathcal{D}, \mathcal{D})\in\mathbb{R}^{n\times n}$ is the prior covariance matrix. 
The estimate of the function $f$ at any unobserved design point $\bx^\ast$ is obtained using the following posterior distribution
\begin{align}
    &P(\hat{f}(\bx^\ast)|f(\mathcal{D}),\theta) = \mathcal{N}(\mu_{\text{pos}}(\bx^\ast), \sigma_{\text{pos}}^2(\bx^\ast)),
 \end{align}
 where, with $k_{\theta}(\bx^\ast,\mathcal{D})\in\mathbb{R}^{1\times n}$, $k_{\theta}(\mathcal{D},\bx^\ast)\in\mathbb{R}^{n\times 1}$, presuming additive Gaussian independent identically distributed data noise with zero mean and $\sigma^2_{\text{noise}}$ variance, and a prior GP mean function $\mu(\bx)\equiv \bm{0}$, we have 
 
 \begin{align}
 \begin{split}
    &\mu_{\text{pos}}(\bx^\ast) =  k_{\theta}(\bx^\ast, \mathcal{D}) (k_{\theta}(\mathcal{D}, \mathcal{D})+\sigma^2_\mathrm{noise}I_n)^{-1}f(\mathcal{D})  \\
   &\sigma_{\text{pos}}^2(\bx^\ast) = k_{\theta}(\bx^\ast,\bx^\ast)- \\
    & \quad k_{\theta}(\bx^\ast, \mathcal{D}) (k_{\theta}(\mathcal{D}, \mathcal{D}) +\sigma^2_\mathrm{noise}I_n)^{-1}  k_{\theta}(\mathcal{D}, \bx^\ast)
 \label{eq:posterior}
 \end{split}
\end{align}

The $\mu_{\text{pos}}(\bx^\ast)$ and $\sigma_{\text{pos}}^2(\bx^\ast)$ are the posterior mean and variance of the function $f$ at $\bx^\ast$ conditioned on the observed data, respectively. 
We employ an anisotropic squared exponential RBF kernel, where, for any two points $(\bx,\bx')$,  
\begin{align}
k_\theta(\bx, \bx') &= A \cdot \exp\left(-(\bx - \bx')^T M^{-1}(\bx - \bx')\right)
\label{eq:kernel}
\end{align}
where $M = \text{diag}(l_\text{po}, l_\text{pr})\in\mathbb{R}^{2}$, and $(l_\text{po}, l_\text{pr})$ are the correlation lengths in the power and pressure directions, respectively.
The values of hyperparamters $\theta = \{A, l_\text{po}, l_\text{pr}\}$ are selected using hyperparamter tuning, discussed in \cref{sec:hyper_tune}.
Finally, the posterior mean and variance at all points in search space $\mathcal{X}$ is used to estimate the acquisition function; see \cref{sec:ucb}.

\subsubsection{Selecting Hyperparameters of Gaussian Process}\label{sec:hyper_tune}
The hyperparameters $\theta$ used in the Gaussian process for building the surrogate model during Bayesian optimization and the hyperparameters $\phi$ for the interpolant $g(\bx)$ are selected by maximizing the log-likelihood of the observed data $f(D)$, as shown below with $\lambda:=(\theta,\sigma_\text{noise})$
\begin{align}
\hat{\theta} &= \argmax_{\theta} \log P(f(\mathcal{D})|\lambda) \label{eq:mle}\\
\log P(f(\mathcal{D})|\lambda) &= - \frac{1}{2}f(\mathcal{D})^T (k_{\theta}(\mathcal{D},\mathcal{D}) + \sigma_\text{noise}^2I_n)^{-1}f(\mathcal{D}) \notag\\
&\quad -\frac{1}{2} \log |k_{\theta}(\mathcal{D},\mathcal{D}) + \sigma_\text{noise}^2I_n| \notag \\
&\quad - \frac{n}{2}\log{2\pi}. \label{eq:lle}
\end{align}
The optimization problem, \cref{eq:mle}, is solved using gradient descent on \cref{eq:lle}. 
Since the unified objective function $f(\cdot)$ is deterministic, we choose a predetermined small value of $\sigma_\text{noise}=0.01$ for regularization purposes. As for the GP regression context for the stress surrogate in Eq.~({\ref{eq:interpolant}}) above, we used $\sigma_\text{noise}=0.058$ based on the observed data noise from the experiments conducted prior to BayesOpt.

\subsubsection{Acquisition Function} \label{sec:ucb}
The acquisition function provides the means for BayesOpt to balance exploration and exploitation. We use the upper confidence bound (UCB) as the acquisition function to sample new power and pressure values where the sputter deposition is conducted. We chose UCB given its simplicity and broad utility. The UCB acquisition function characterizes the input space $\mathcal{X}$ into regions for exploration with high uncertainty and regions for exploitation that likely contain the objective function's optimum. 
The UCB function is the sum of the posterior mean and scaled standard deviation obtained from the surrogate model, given by
\begin{align}
	\ucb(\bx) &= \mu_{\text{pos}}(\bx) + \beta \cdot \sigma_{\text{pos}}(\bx), \label{eq:ucb}
\end{align}
where the hyperparameter $\beta$ controls the balance between exploration and exploitation. 
A larger value of $\beta$ will drive the BayesOpt algorithm to enhance its priority for exploration over exploitation. The solution of the UCB optimization problem is the next design point in the search space $\mathcal{X}$, given by
\begin{align}
	\bar{\bx} &= \argmax_{\bx \in \mathcal{X}} \ucb(\bx). \label{eq:next_point}
\end{align}
The optimal solution of the UCB function is obtained using gradient-based optimizer L-BFGS-B with several random starts to estimate good hyperparameter values. 
The optimum value $\bar{\bx}$ is the next point where subsequent sputter deposition is conducted. 

\subsection{Algorithm termination strategy}\label{sec:terminate}

Our approach for algorithm termination hinges on achieving the convergence of the optimal solution $\bx^*$.
During BayesOpt guided search, the optimal solution at each iteration is identified as the configuration within the current dataset $\mathcal{D}$ that maximizes the objective function $f(x)$ as shown below.
\begin{align}
 \bx^* = \argmax_{ \bx \in \mathcal{D}} f(\bx) \label{eq:optimum}
\end{align}
The optimal solution is determined by simply identifying the configuration for which the function attains its maximum value in the set $f(\mathcal{D})$.
We monitor the optimal solution at each iteration until its value remains stagnated for a maximum predefined count. 
Subsequently, we assess all four objective criteria, objective function, and uncertainty and conclude the termination of the experiment. 

\section{Experimental Setup}\label{sec:Ex_setup}

\begin{figure}[htbp]
  \begin{subfigure}[l]{19pc}
    \includegraphics[width=\textwidth]{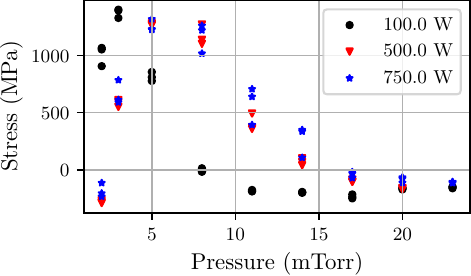}
    \caption{}
    \label{fig:exp_stress}
  \end{subfigure}
  \begin{subfigure}[l]{19pc}
  \vspace{2em}
  \includegraphics[width=\textwidth]{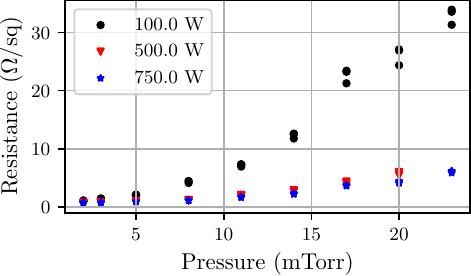}
    \caption{}
    \label{fig:exp_resist}
  \end{subfigure}
  \caption{
The plots depict the variations of residual stress, shown in \cref{fig:exp_stress}, and variations of sheet resistance, shown in \cref{fig:exp_resist} with pressure and powers in the dataset before Bayesian optimization. 
The observations from the plots assisted in constructing the objective functions. 
}
\label{fig:stress_resistance_observed}
\end{figure}

For this work, the stress and sheet resistance threshold values were chosen as $T_s=300$ MPa and $T_r=3$ $\runit$. 
The BayesOpt-guided experiment was started with the initial measurements of stress and sheet resistance of the thin film, deposited using $27$ different configurations of power and pressure settings, as shown in \cref{fig:stress_resistance_observed}. 

At each iteration, a pair of power and pressure settings for the sputter deposition of Mo films was recommended by the BayesOpt algorithm. 
Subsequently, the deposited films' stress and sheet resistance were measured and fed back into the BayesOpt algorithm for the next iteration.
\cref{sec:bayes_run} describes the settings under which the Bayesian optimization algorithm was executed.
\cref{sec:deposition} describes the sputter deposition of Mo films, and sections \ref{sec:stress_measurement}  and \ref{sec:resistance_measurments} describe the measurements of stress and sheet resistance, respectively.

\subsection{Bayesian optimization and objective function setup}\label{sec:bayes_run}

The BayesOpt algorithm was initialized using stress and sheet resistance measurements for the $27$ initial configurations. 
The BayesOpt search space, $\mathcal{X}$, was defined between the pressure values of $2$ mTorr to $23$ mTorr and the power values ranging from $50$ W to $750$ W.
The objective function constants were set as $m=5$ in \cref{eq:resistance_criteria} and $B=1900$ in \cref{eq:min_grad_criteria}.
The Gaussian process hyperparameters $\phi$, for the interpolant $g(\bx)$ and $\theta$ for the surrogate model, were tuned as explained in \cref{sec:hyper_tune}.
Finally, $\beta=1$ was used for the UCB function.

The code for the BayesOpt algorithm was implemented in Python 3.9.2 and executed on a Ubuntu 20.04.6 LTS 64-bit machine with 32 Gb RAM and 11th Gen Intel Core i7-1185G7 @ 3.00GHz × 8 processor. 
During each iteration, the algorithm required roughly $10-12$ minutes to finish the computation and provide the next sputter deposition configuration.

\subsection{Thin Film Deposition}\label{sec:deposition}

\begin{figure}[htbp]
    \centering
  \begin{subfigure}{20.25pc}
    \includegraphics[width=\textwidth]{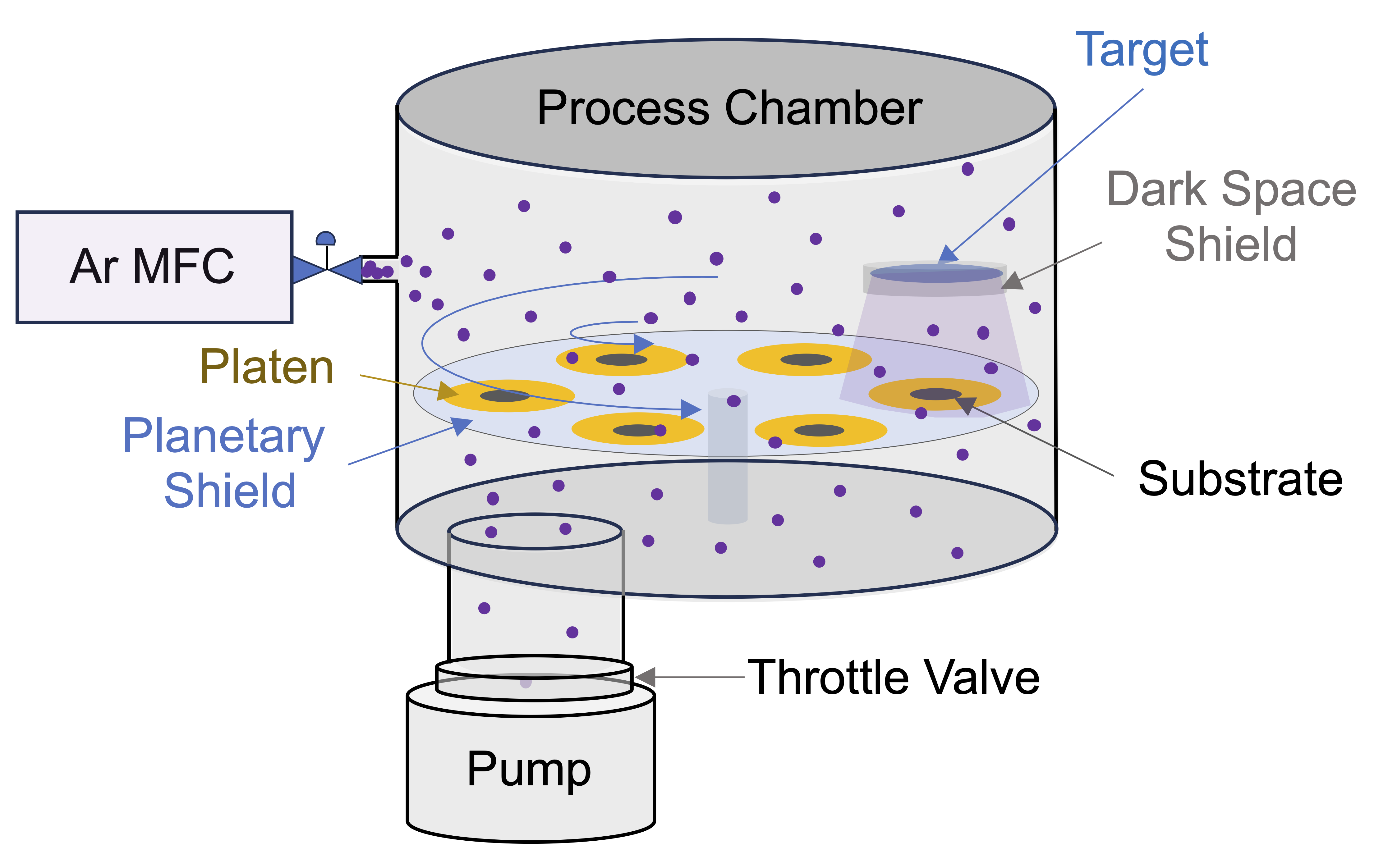}
    \caption{Process Chamber}
    \label{fig:chamber}
  \end{subfigure}
  \begin{subfigure}{20.25pc}
    \includegraphics[width=\textwidth]{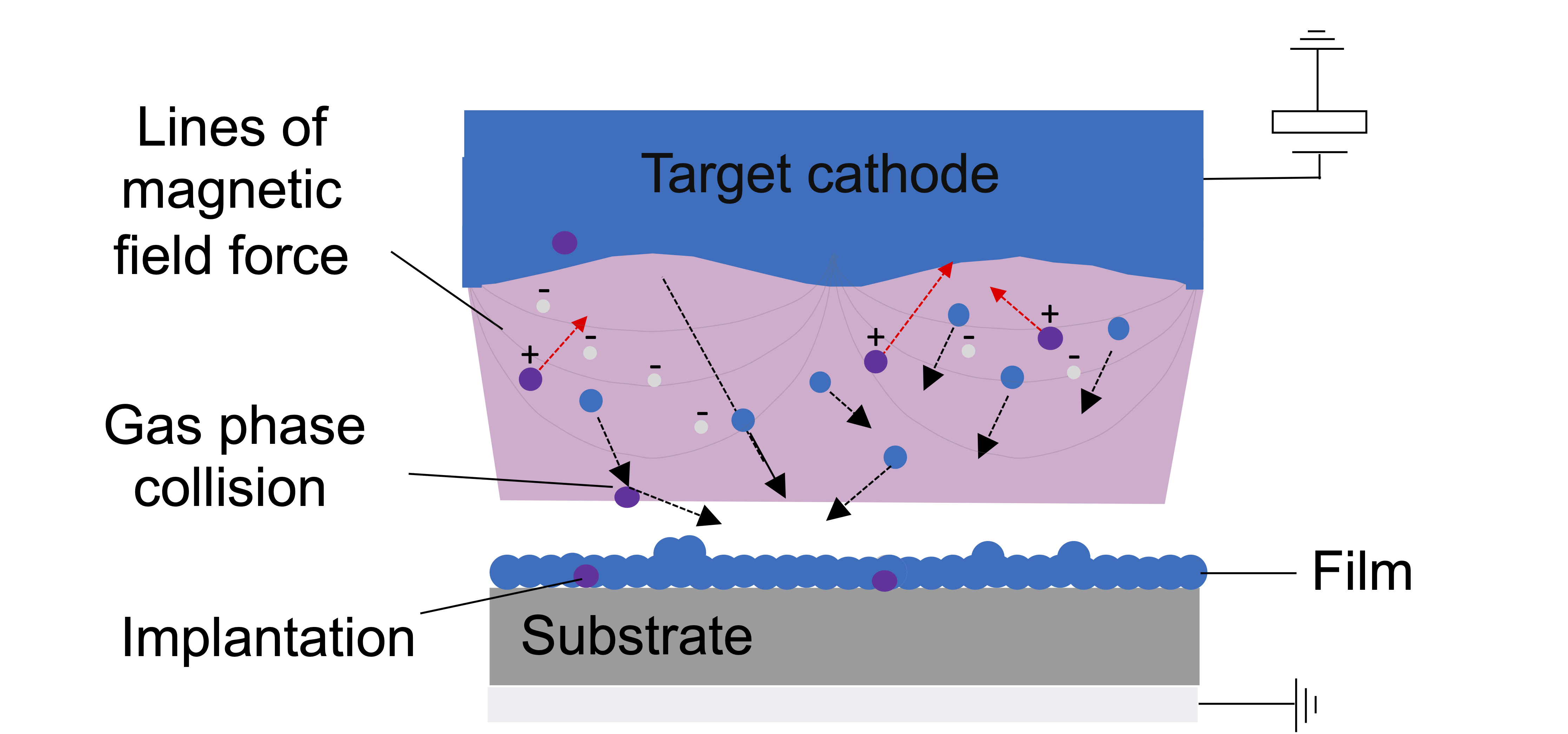}
    \caption{Sputtering Process}
    \label{fig:deposition}
  \end{subfigure}
  \caption{
  Vacuum chamber geometry, as shown in \cref{fig:chamber}, is used for physical vapor deposition employing a top-down sputter geometry with true planetary sample stage motion (i.e., orbit + spin).
  Ultra-high-purity argon gas is consistently introduced into the chamber and regulated by a mass flow controller (MFC).
  \cref{fig:deposition} shows the cross-section drawing of key atomistic processes involved in sputtering, transport, and film growth.}
  \label{fig:thin_film_fabrication}
\end{figure}

Molybdenum films were sputter deposited in an Innotec $240$ SD system, as illustrated in \cref{fig:thin_film_fabrication}. 
Before deposition, the chamber was cryo-pumped to a base pressure of $\le 5 \times 10^{-7}$ Torr. 
A planar target $20.32$ cm ($8$ inches) in diameter and made of high-purity Mo $(99.95\%)$ was used to deposit the material in a sputter-down configuration. 
The substrate rotation stage consists of six platens moving in planetary motion, with a working distance of $\approx$9.25~cm between the substrates and target. 
All films were deposited onto $5.08$~cm ($2$ inches) diameter Si ($100$) substrates with $400$~nm of thermal oxide. 
These substrates were chosen because they possess a single biaxial modulus, significantly simplifying internal stress value calculations using Stoney's equation. 
Three nominally identical Si wafers were loaded onto three platens for redundancy for each deposition.
Ultra-high purity argon flowed into the chamber at $50$ sccm, and its working pressure was maintained via a downstream pressure control (throttled gate valve). 
Ar pressure was held constant and monitored to ensure any changes in pressure were $< 1\%$ for each deposition. 
The same deposition time was used for all sputter pressures so that the samples spent similar time in the flux of the sputter gun. 

\subsection{Substrate and Film Thickness}
Substrate thickness was measured at five locations on each wafer using a drop gauge. 
Each deposition included an additional Si piece of $\approx~1$~sq.~cm for film thickness measurements. 
A portion of this ride-along sample was masked with Kapton tape so that its removal after deposition would create a step edge amenable to thickness measurements. 
To maintain congruity with other Si substrates, witness pieces were of similar thickness, placed in the center of a deposition platen at the same working distance, and exposed to the flux for the same time as other samples. 
A DEKTAK XT surface profilometer measured the film height at five step-edge locations, from which an average film thickness was obtained. 

\subsection{Stress Measurement of Films}\label{sec:stress_measurement}
Before the Mo deposition, an FLX2320-S wafer scanning instrument, which measures the deflection from an incident laser beam, was used to determine an initial radius of curvature for each wafer. 
After deposition, each coated wafer was re-measured with the instrument to find the post-deposition wafer radius of curvature. 
Using these values – together with the film thickness, the wafer thickness measured previously, and the biaxial modulus of Si (100) – the average in-plane stress of the film was calculated according to Stoney's equation~\cite{stoney1909tension}. 

\subsection{Sheet Resistance Measurement of Films}\label{sec:resistance_measurments}
The Mo film's room temperature sheet resistance measurements were made using a four-point probe and a constant current source and meter. 
Samples were probed with a single tip assembly having four $100~\mu$m radius tips arranged with 1.0 mm spacing. 
Sheet resistance was measured six times for each of the Mo-coated wafers at three locations relative to the wafer flat (top, bottom, and center) and in forward bias and reverse bias at each position. 
Insulating substrates (i.e., with thermal oxide) helped isolate the measurement to be only of the film. 

\section{Results}\label{sec:results}
The BayesOpt-guided sputter deposition experiment continued for $10$ iterations until the change in optimal configuration $\bx^* = (2 \text{ mTorr}, 620 \text{ W})$ no longer exceeded the predefined thresholds in both power ($\epsilon_{\text{pr}}=0.5$) and pressure ($\varepsilon_{\text{po}}=10$).

During each iteration of the experiments, multiple thin films were deposited using the configuration proposed by BayesOpt, denoted as $\bar{\bx}$, and their properties were measured. 
Subsequently, these measured properties were used to update the objective function $f(x)$ and estimate the optimal configuration, $\bx^*$.
For additional information about the experiments and property measurements, please refer to Appendix {\ref{sec:appendix}}.

The trends of the BayesOpt algorithm proposals are discussed in Section { \ref{sec:trend}}, and the rationale behind the conclusion about the optimal solution is detailed in Section {\ref{sec:verification}}.
For the convenience of the discussion, we denote the subregion of the search space $\mathcal{X}$, with pressure $<3$~mTorr as $\mathcal{X}_1$ and that with pressure $\geq 3$~mTorr as $\mathcal{X}_2$.

\subsection{Bayesian Optimization iteration trends}\label{sec:trend}

\begin{figure}[htbp]
	\centering
  \begin{subfigure}{19pc}
    \includegraphics[width=\textwidth]{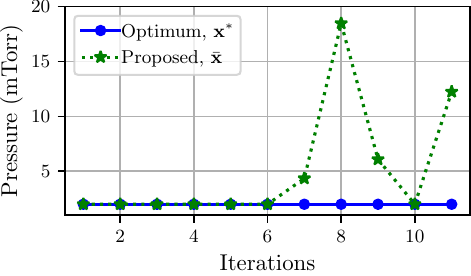}
    \caption{Pressure v/s iterations}
    \label{fig:iter_pressure}
  \end{subfigure}

  \begin{subfigure}{19pc}
    \vspace{1em}
    \includegraphics[width=\textwidth]{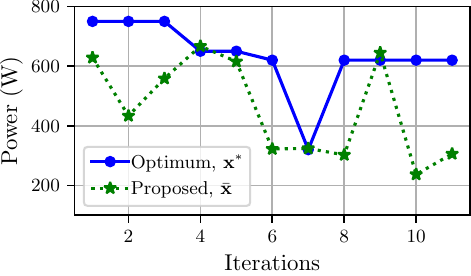}
    \caption{Power v/s iterations}
    \label{fig:iter_power}
  \end{subfigure}
  \caption{
    The figures illustrate trends observed during the BayesOpt guided search, including the proposed sputter deposition configuration for the next iteration and the optimal configuration in that iteration.
    Figs.~\ref{fig:iter_pressure}  and \ref{fig:iter_power} show pressure and power trends over iterations, respectively.
    See \cref{tab:trends_deposition} in \cref{sec:appendix} for the specific values.
  }
  \label{fig:config_trends}
\end{figure}

\begin{figure}[htbp]
 \begin{subfigure}{19pc}
    \includegraphics[width=\textwidth]{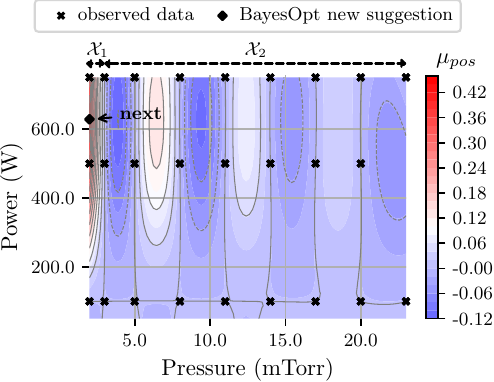}
    \caption{Exploitation term, $\mu_\text{pos}(\bx)$}
    \label{fig:mu0}
\end{subfigure}
 \begin{subfigure}{19pc}
    \vspace{1em}
    \includegraphics[width=\textwidth]{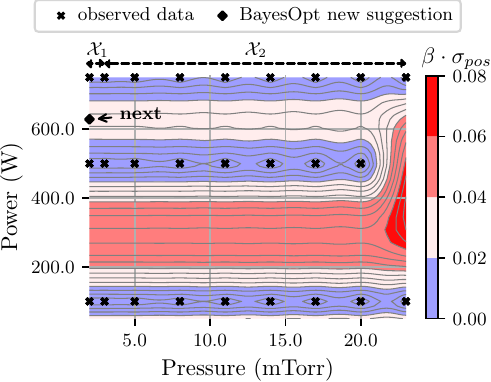}
    \caption{Exploration term, $\beta \cdot \sigma_\text{pos}(\bx)$}
    \label{fig:sigm0}
\end{subfigure}
 \begin{subfigure}{19pc}
 \vspace{1em}
    \includegraphics[width=\textwidth]{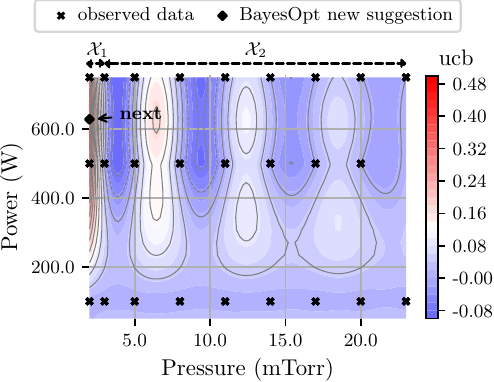}
    \caption{Acqusition function, $\ucb(\bx)$}
    \label{fig:only_ucb0}
\end{subfigure}
\caption{The figure shows the contour plots of exploitation term, $\mu_\text{pos}$ in \cref{fig:mu0}, exploration term $\beta \cdot \sigma_\text{pos}$ in \cref{fig:sigm0}, and UCB in \cref{fig:only_ucb0} for the first iteration with $\beta=1$. 
At this iteration, BayesOpt is exploiting the region $\mathcal{X}_1$.} 
\label{fig:ucb0}
\end{figure}

\begin{figure}[htbp]
 \begin{subfigure}{19pc}
    \includegraphics[width=\textwidth]{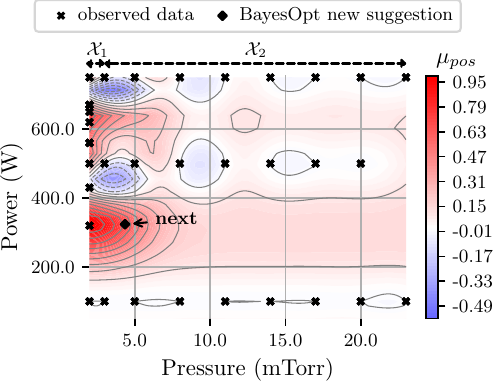}
    \caption{Exploitation term, $\mu_\text{pos}(\bx)$}
    \label{fig:mu7}
\end{subfigure}
 \begin{subfigure}{19pc}
    \vspace{1em}
    \includegraphics[width=\textwidth]{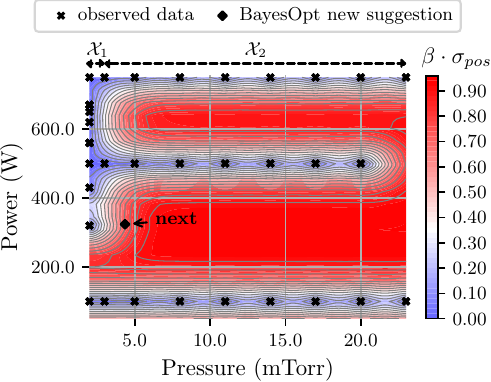}
    \caption{Exploration term, $\beta \cdot \sigma_\text{pos}(\bx)$}
    \label{fig:sigm7}
\end{subfigure}
 \begin{subfigure}{19pc}
    \vspace{1em}
    \includegraphics[width=\textwidth]{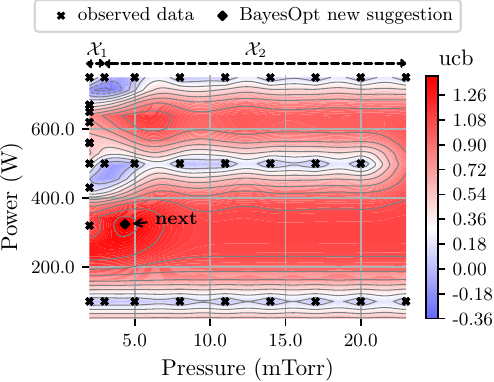}
    \caption{Acqusition function, $\ucb(\bx)$}
    \label{fig:only_ucb7}
\end{subfigure}
\caption{The figure shows the contour plots of exploitation term, $\mu_\text{pos}$ in \cref{fig:mu7}, exploration term $\beta \cdot \sigma_\text{pos}$ in \cref{fig:sigm7}, and UCB in \cref{fig:only_ucb7} for the seventh iteration with $\beta=1$. 
At this iteration, BayesOpt is exploring the region $\mathcal{X}_2$.} 
\label{fig:ucb7}
\end{figure}

\begin{figure}[htbp]
 \begin{subfigure}{19pc}
    \includegraphics[width=\textwidth]{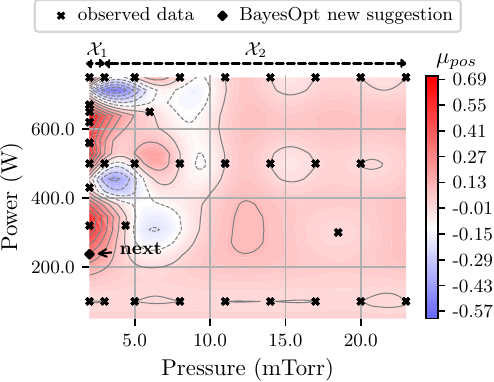}
    \caption{Exploitation term, $\mu_\text{pos}(\bx)$}
    \label{fig:mu10}
\end{subfigure}
 \begin{subfigure}{19pc}
   \vspace{1em}
    \includegraphics[width=\textwidth]{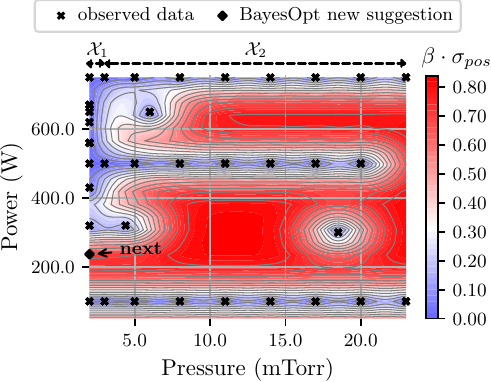}
    \caption{Exploration term, $\beta \cdot \sigma_\text{pos}(\bx)$}
    \label{fig:sigm10}
\end{subfigure}
 \begin{subfigure}{19pc}
   \vspace{1em}
    \includegraphics[width=\textwidth]{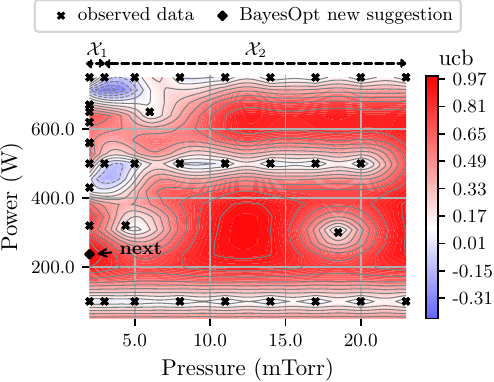}
    \caption{Acqusition function, $\ucb(\bx)$}
    \label{fig:only_ucb10}
\end{subfigure}
\caption{The figure shows the contour plots of exploitation term, $\mu_\text{pos}$ in \cref{fig:mu10}, exploration term $\beta \cdot \sigma_\text{pos}$ in \cref{fig:sigm10}, and UCB in \cref{fig:only_ucb10} for the tenth iteration with $\beta=1$. 
At this iteration, BayesOpt is exploring the region $\mathcal{X}_2$.} 
\label{fig:ucb10}
\end{figure}

The trends in the configurations proposed by BayesOpt, denoted as $\bar{\bx}$, for the next sputter deposition experiment at each iteration are shown in \cref{fig:config_trends}.
It was observed that for the first six iterations, BayesOpt proposed configurations with a fixed pressure value of $2$ mTorr and varying power (region $\mathcal{X}_1$). 
Then, in later iterations, it proposed configurations in the complementary region, $\mathcal{X}_2$.
To offer insight into the rationale behind the suggested trends, the contour plots of the posterior mean $\mu_{\text{pos}}(\bx)$, standard deviation $\sigma_{\text{pos}}(\bx)$ and $\ucb(\bx)$ at the first, seventh, and tenth iterations are shown in Figs.~\ref{fig:ucb0}, \ref{fig:ucb7} and \ref{fig:ucb10} respectively.

During the first six iterations, as shown \emph{e.g.} for the first iteration in \cref{fig:ucb0}, contour plots show that the posterior mean, $\mu_{\text{pos}}(\bx)$ in $\mathcal{X}_1$ dominated the standard deviation term, $\beta \cdot \sigma_{\text{pos}}(\bx)$ in the entire search space in terms of influencing $\ucb(\bx)$. 
The high posterior mean in $\mathcal{X}_1$ was due to high values in the first three design criteria, $f_1, f_2, f_3$, suggesting a likely location for the objective function maxima.
Accordingly, the algorithm exploited the region $\mathcal{X}_1$ to find the optimum. 
Later, during iterations $7$, $8$ and $9$, as shown \emph{e.g.} for the seventh iteration in \cref{fig:ucb7}, contour plots show the standard deviation, $\beta \cdot \sigma_{\text{pos}}(\bx)$ in the region $\mathcal{X}_2$ dominating the posterior mean in terms of influencing $\ucb(\bx)$. 
Hence, the algorithm explored the region $\mathcal{X}_2$ due to the lack of observations causing high local uncertainty.  
In the $10$th iteration, as shown in \cref{fig:ucb10}, the algorithm revisited exploitation within the $\mathcal{X}_1$ region, only to shift back to exploration in the $\mathcal{X}_2$ region during the $11$th iteration.
In summary, the algorithm goes back and forth between exploration and exploitation to efficiently search for the optimal solution by balancing the need to gather information about the objective function and leveraging that information to refine the search.

During the experiments, we evaluated the optimal solution $\bx^*$ at each iteration, as shown in \cref{fig:config_trends}.
It was observed that the optimal solution changed during the initial iterations until it remained consistent at $(2 \text{ mTorr}, 620 \text{ W})$ for the last four iterations.
The Bayesian optimization algorithm was terminated based on the convergence of the optimal solution $\bx^*$.

Note that the final/converged optimal values of stress and resistance, $(2 \text{ mTorr}, 620 \text{ W})$ were in fact proposed at iteration $5$, but the ``optimal`` solution identified by our iterative method remained consistent only after iteration $8$.
This delay occurred because our algorithm also attempts to learn accurate values of the objective function $f$ at the observed configurations $\mathcal{D}$.
The evaluation of the objective function depends not only on the values of stress and resistance but also on the derivative of stress with pressure, refer Eq.~{\ref{eq:combined}}.
Since we cannot directly obtain the derivative values from the experiments, we rely on the interpolant, $g(\bx)$ for calculations.
As more experiments are conducted, $g(\bx)$ converges to a more accurate representation of the stress function $S(\bx)$ in the region $\mathcal{X}_1$, as does its derivative.
This leads to the convergence of our estimated objective function values $f(\mathcal{D})$.
Thus, the delayed convergence of the optimal solution highlights the algorithm's process of learning and adapting to the characteristics of the objective function, ultimately leading to more accurate estimates over time.

\subsection{Validating the optimal solution}\label{sec:verification}

\begin{figure*}[htbp]
	\centering
  \begin{subfigure}{20.25pc}
    \includegraphics[width=\textwidth]{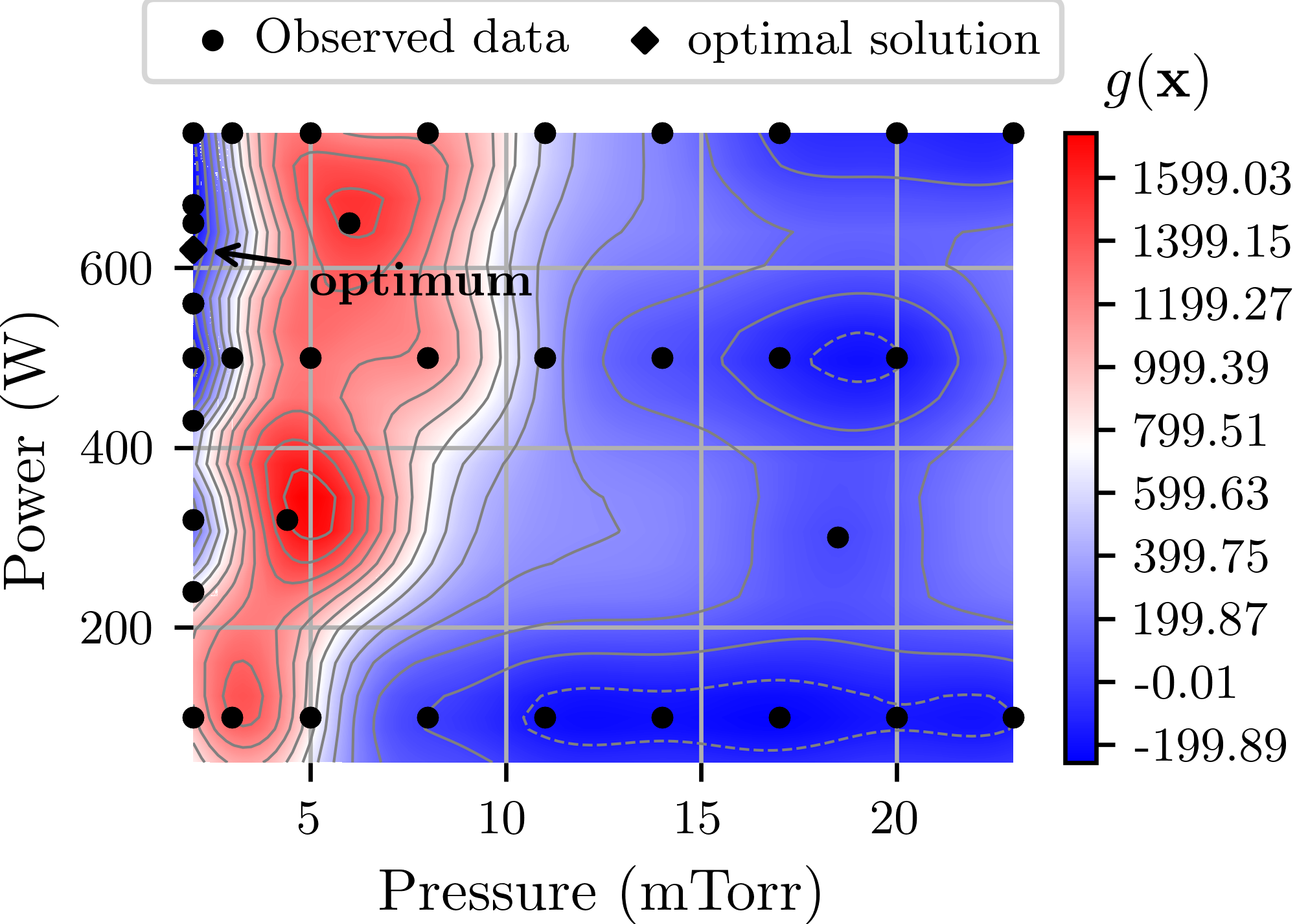}
    \caption{residual stress interpolant, $g(\bx)$ in MPa}
    \label{fig:sub1}
  \end{subfigure}
  \hspace{1pc}
  \begin{subfigure}{20.25pc}
    \includegraphics[width=\textwidth]{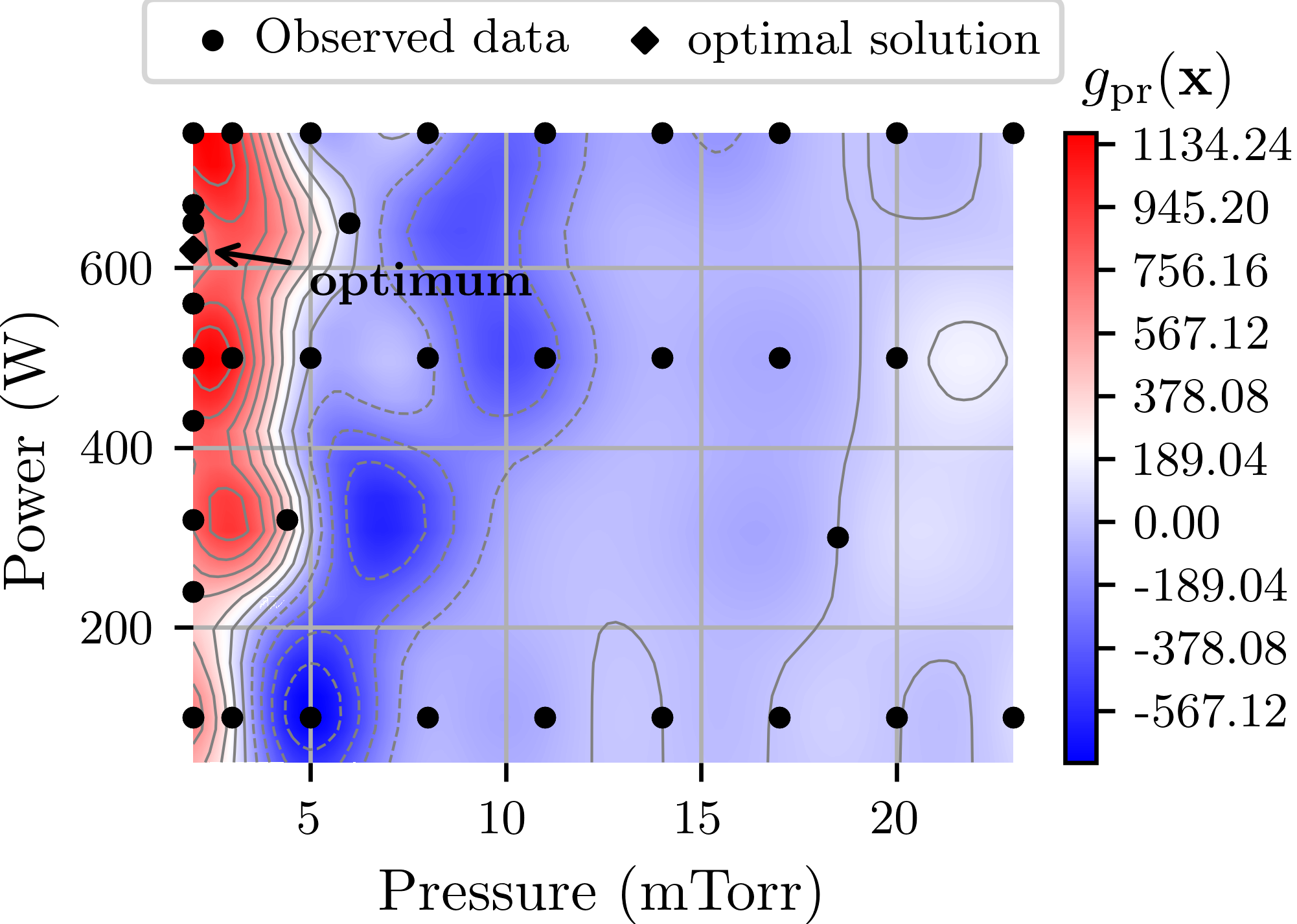}
    \caption{derivative of stress with pressure, $g_{\text{pr}}(\bx)$}
    \label{fig:sub3}
  \end{subfigure}

  \begin{subfigure}{20.25pc}
    \vspace{2em}
    \includegraphics[width=\textwidth]{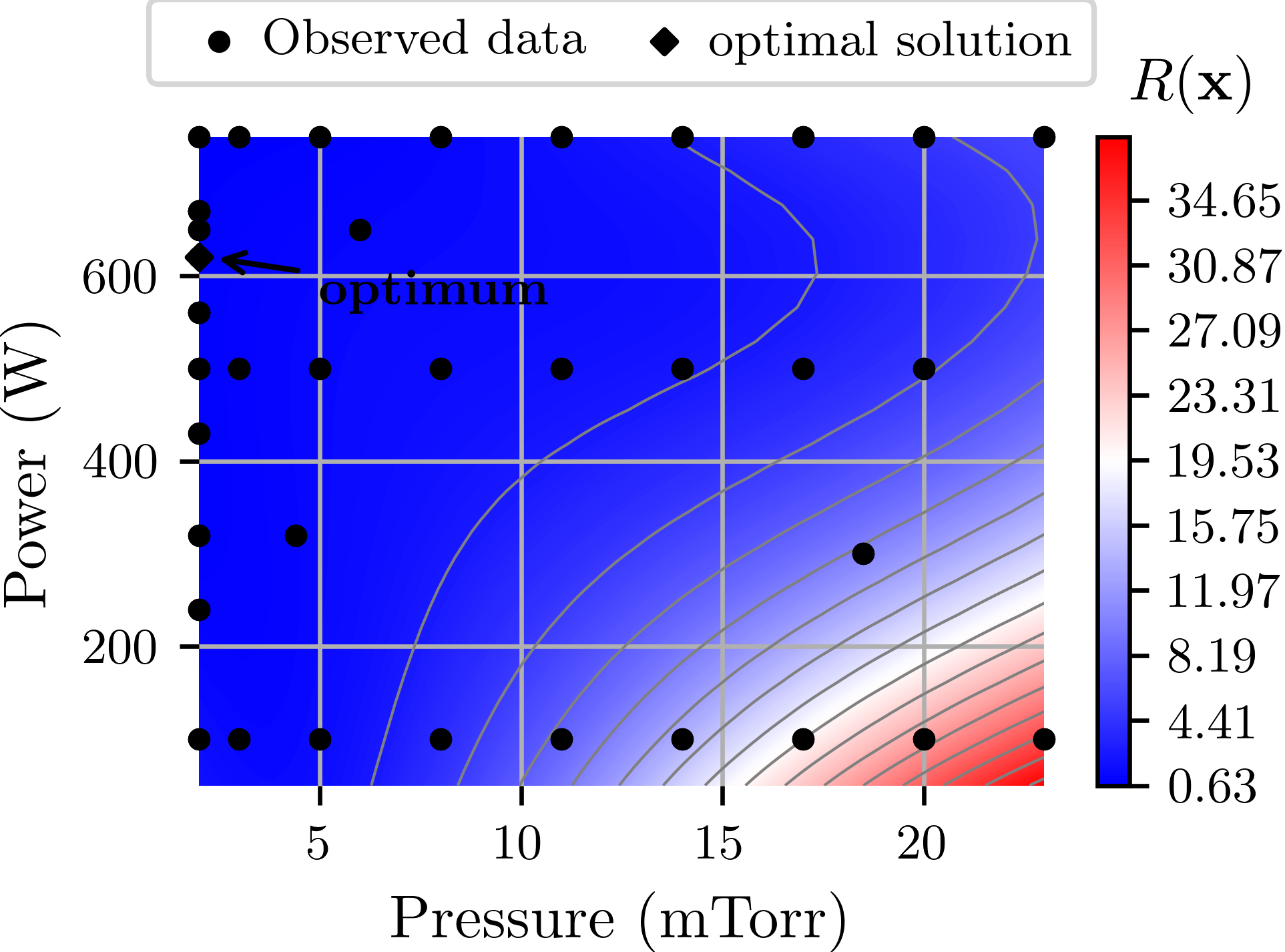}
    \caption{sheet resistance, $R(\bx)$ in $\runit$}
    \label{fig:sub2}
  \end{subfigure}
  \hspace{1pc}
  \begin{subfigure}{20.25pc}
   \includegraphics[width=\textwidth]{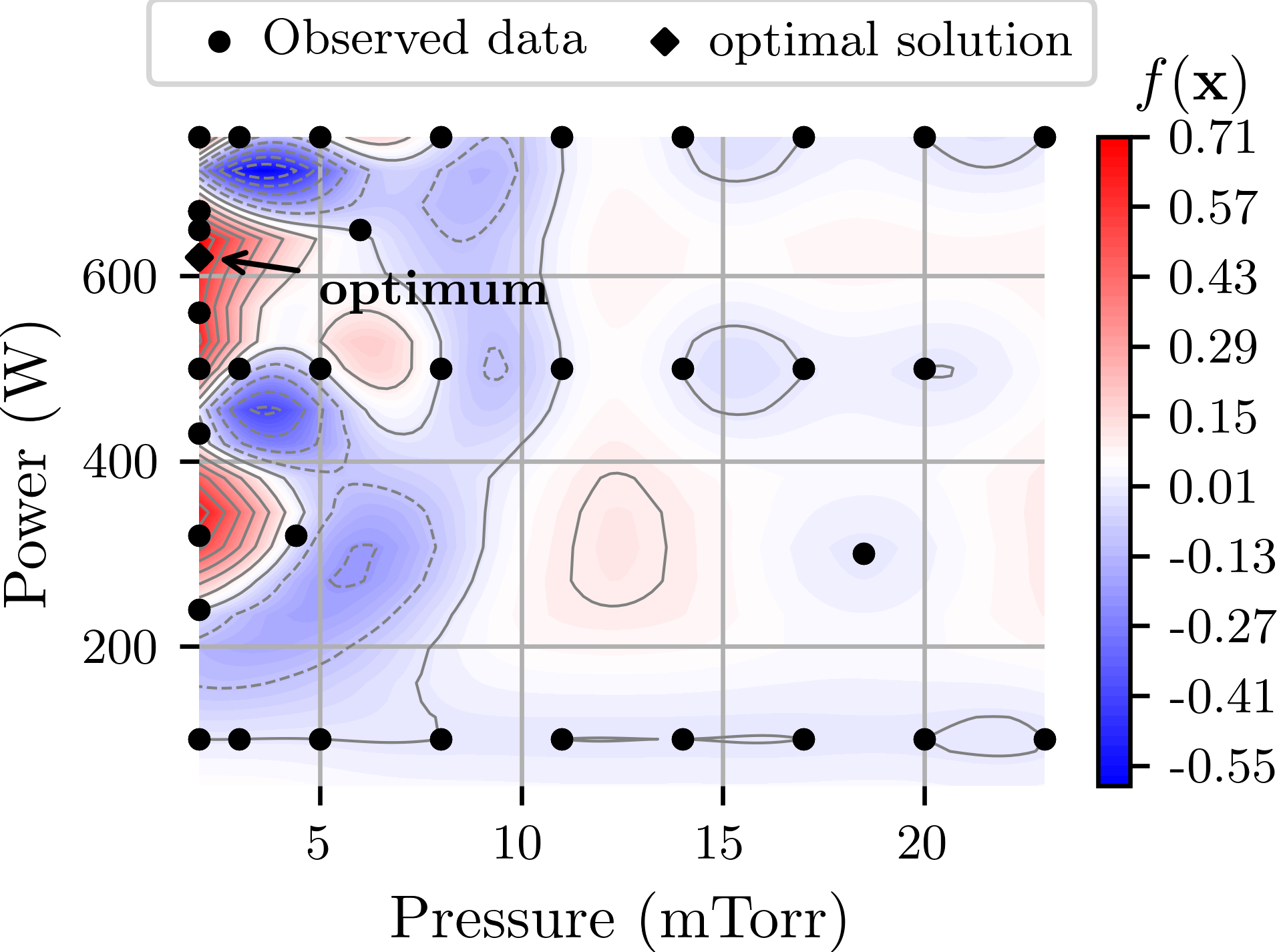}
    \caption{objective function, $f(\bx)$
    \label{fig:sub4}
    }
    \label{fig:final_obj10}
  \end{subfigure}
  \caption{The figure shows the contour plots of stress in \cref{fig:sub1}, the derivative of stress with pressure in \cref{fig:sub3}, the sheet resistance in \cref{fig:sub2}, and the objective function $f$ in \cref{fig:sub4} after $10$ iterations. 
  The optimal conditions are achieved at $(2\text{ mTorr}, 620 \text{ W})$, ensuring that the observed stress falls within the range of $[-300, 300]$ MPa, the sheet resistance is below $3$ $\runit$, and the derivative is both positive and at its minimum, fulfilling all specified criteria.
  \label{fig:obj10_2D}
  }
\end{figure*}

Based on the observations in \cref{fig:config_trends} and termination criterion discussed in \cref{sec:terminate}, the optimal configuration was concluded as $(2 \text{ mTorr}, 620 \text{ W})$.
For verification that the optimal configuration for the sputter deposition satisfied the design objectives, we analyzed the sheet resistance values, the interpolant $g(\bx)$ for stress, and the derivative $g_\text{pr}(\bx)$ at the observed configurations $\mathcal{D}$. 
These results are illustrated in \cref{fig:obj10_2D}. 
The optimal configuration was first observed in iteration 5, with resulting stresses $(-180.9, -218.0, -215.5)$ MPa and sheet resistances  $(0.68, 0.68, 0.77)$ $\runit$, refer \cref{sec:appendix} for details. 
These values satisfy stress requirements in $[-300,300]$ MPa, and sheet resistance $< 3$ $\runit$. The optimum is indicated in the field plots in \cref{fig:final_obj10}, putting it in the context of the overall variation of the fields of interest. The derivative of stress with pressure is positive at the optimum, and the overall objective function is high.   

\begin{figure}[htbp]
	\centering
  \begin{subfigure}{19pc}
    \includegraphics[width=\textwidth]{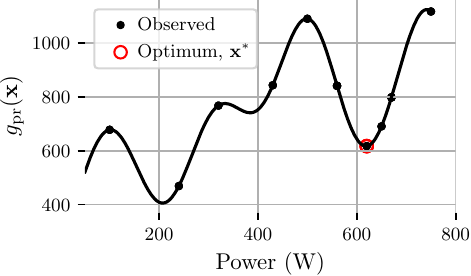}
    \caption{}
    \label{fig:grad10_1D}
  \end{subfigure}

  \begin{subfigure}{19pc}
    \vspace{1em}
    \includegraphics[width=\textwidth]{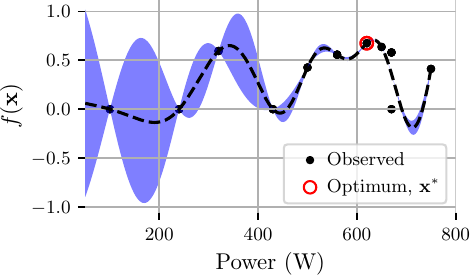}
    \caption{}
    \label{fig:obj10_1D}
  \end{subfigure}
  \caption{
 The plot in \cref{fig:grad10_1D} illustrates the plots of the derivative of stress with pressure, $g_\text{pr}(\bx)$ against power. 
 Similarly, \cref{fig:obj10_1D} illustrates the GP fit of the objective function $f(\bx)$ based on observations at $\mathcal{D}$.
 \cref{fig:obj10_1D} presents the variations of the posterior mean, $\mu_{\text{pos}}(\bx)$, (as a dashed line) and the standard deviation, $\sigma_{\text{pos}}(\bx)$, (as a shaded area) with power.
 with the posterior mean represented  and the standard deviation 
 The peak value of the objective function is identified at the point $(2~\text{mTorr}, 620~\text{W})$. 
 Both plots are shown for the pressure $2$ mTorr, after $10$ iterations.
 }
  \label{fig:1D_10}
\end{figure}

For further clarity, we plot the variation of $g_\mathrm{pr}(\bx)$ and $f(\bx)$ with power, at the optimal pressure of 2~mTorr, in \cref{fig:1D_10}. 
Here, we see, indeed, the optimal power at $620$~W. 
Note that the derivative is, in fact, minimal around $200$ W; however, clearly, the region preceding $300$~W fails to meet the other three criteria, such that full objective function $f(\bx)$ is relatively low in that region. 
Evidently, the configuration with the smallest derivative value that fulfills all criteria is found at $620$~W, exhibiting notably low uncertainty given the locally high data density. 
We note that the other cluster with high $f(\bx)$, although not as high as the optimum, is the region around $(2 \text{ mTorr}, 320 \text{ W})$. 
It is possible that this region, with further exploration, might be competitive with the optimum. 
However, since we stopped at ten iterations, given experimental budget constraints, we have the optimal solution at $\bx^\ast=($2\,mTorr, 620\,W$)$. 

The optimal configuration (2\,mTorr, 620\,W) met the stress and resistance criteria, with stress values of $(-180.9, -218.0, -215.5)$\,MPa and sheet resistances of $(0.68, 0.68, 0.77)$\,$\runit$. 
Further, to ensure that this configuration is least affected by minor pressure fluctuations, additional experiments were conducted in the region $\mathcal{X}_1$ to analyze the sensitivity of film stress to pressure.
We conducted experiments around the optimal power setting of $620$ W, exploring values at $590$ W and $650$ W, each differing by $30$ W from the optimal setting. 
Additionally, we varied the pressure incrementally by $0.5$ mTorr, starting at the optimal pressure of $2$ mTorr.
The observations of mean stress for this experimental setup are shown in Table {\ref{tab:additional_exp}}. More detailed information on these additional experiments is provided in Appendix {\ref{sec:appendix_additional}}.
For sensitivity analysis, we computed the derivative of stress with pressure corresponding to each power setting, as presented in the table. 
Results revealed that stress exhibited the least sensitivity to pressure at $(2$\,mTorr, 620\,W$)$, with a derivative value of $339.23$, which increased away from the optimal conditions in either power or pressure while satisfying all the other criteria, further supporting the conclusions drawn from the BayesOpt algorithm.

\begin{table}[htp]
    \centering
    \caption{Table displaying the mean of residual stresses in thin films deposited using the utilized configuration $\bx$ during supplementary experiments.
    The sensitivity of stress to pressure was analyzed by computing the derivative of stress with respect to pressure for each power setting.
    The derivative is smallest at $(2 \text{mTorr}, 620 \text{W})$, further supporting the optimum observed in \cref{fig:1D_10}. 
    }
    \begin{tabular}{C{6pc}C{6pc}C{6pc}}
    \hline \hline
    utilized configuration & mean stress (MPa) & derivative of stress with pressure \\ 
     $\bx$ & $\bar{S}$ & {\large $\frac{\Delta \bar{S}}{\Delta x_{\text{pr}}}$} \\
    \hline \hline
    $(2, 590)$ & $-85.58$ & \multirow{2}{*}{$975.5$}\\
    $(2.5, 590)$ & $402.17$  &\\
    \hline
    $(2, 620)$ & $-44.00$  & \multirow{2}{*}{$399.23$}\\
    $(2.5 , 620)$ &	$155.62$  &\multirow{2}{*}{$1030.16$}\\
    $(3 , 620)$ &	$670.70$  &\\
    \hline
    $(2, 650)$ & $-256.62$ & \multirow{2}{*}{$1115.6$}\\
    $(2.5, 650)$ & $301.06$ & \\
    \hline \hline
    \end{tabular}
    \label{tab:additional_exp}
\end{table}

\section{Conclusion}\label{sec:conclusion}

We used Bayesian optimization to identify optimal operating conditions for the sputter deposition of Mo thin films on Si, targeting desired robust film properties. 
Specifically, the objective was to find a combination of power and pressure such that film stress is in the range of $(-300, 300)$~MPa, the sheet resistance is below $3$~mTorr, and the film is dense. 
We used a positive derivative of stress vs. pressure requirement to identify sputter configurations that lead to dense films. 
Although density was not directly measured in this work, we reasonably conclude that the positive stress gradient requirement promoted denser films. 
This is supported by our previous studies of Mo films using the same apparatus described herein, which show by Rutherford Backscattering Spectrometry that lower process pressures (near the positive stress gradient) produce denser films\cite{kalaswad2023sputter}. 
Further, to promote stability in stress against pressure fluctuation, we targeted conditions with a low gradient of stress vs. pressure. 
Four distinct smooth functions were employed to achieve these objectives, using a unified objective function obtained by multiplying the individual objective functions.  

The Bayesian optimization algorithm performs an adaptive search over the parameter space for optimal solutions, iteratively proposing the next sputter configuration for thin film fabrication and subsequent stress and sheet resistance measurement. 
In each iteration, stress values for thin films obtained from known configurations were utilized to construct a Gaussian process mean interpolant, whose derivative can be evaluated analytically. 
The objective function values at the known configurations were derived from sheet resistance, stress, and the derivative of stress with respect to pressure. 
A Gaussian process was fitted to these objective function values, providing predictive posterior mean and variance. 
These, in turn, were used to compute the upper confidence bound function, guiding the algorithm to propose the next sputter deposition configuration. 
The Bayesian optimization guided search continued for $10$ iterations until the calculated optimal solution at each step remained consistent.
It was observed that the sputter configuration that maximized the objective function was ($2$ mTorr, $620$ W).

The value of stress, sheet resistance, and derivative of stress with pressure were verified at the optimal solution.
It was observed that all the criteria were satisfied. 
The stress values at the optimal configuration were $(-180.9, -218.0, -215.5)$ MPa, falling within the desired range of $(-300, 300)$ MPa, satisfying the first criteria.
Similarly, the sheet resistance measurements were $(0.68, 0.68, 0.77) \ \runit$ remaining below $3 \ \runit$ which satisfied the second criteria.
Finally, the derivative of stress with pressure at the optimal configuration was positive and sufficiently low, satisfying the last two criteria. 
Moreover, further experiments conducted for sensitivity analysis near the BayesOpt proposed optimum configuration provided supporting evidence that this configuration was indeed at a local optimum.

We found that Bayesian optimization efficiently explored the configuration space, identifying regions likely to yield optimal results.
This work also showed that a unified objective function can simultaneously achieve multiple desired properties of Mo thin films while ensuring their robustness against minor variations in sputter deposition conditions. 
While we did not explore different weighting of the various objectives, \emph{e.g.} where weight exponents $(w_1,\ldots,w_4)$ are used, resulting in $f(\bx):=\prod_{i} f_i^{w_i}(\bx)$, this can easily be done, allowing evaluation of selective weighting strategies. 
Moreover, comparing the convergence and efficiency of constrained optimization using a filter-based objective function with other Bayesian optimization approaches that incorporate constraints would be useful.
The present approach can be extended to more complex problems involving high dimensional multimodal datasets involving \emph{e.g.} X-ray diffraction and scanning electron microscope image data.

\begin{acknowledgments}
The authors acknowledge funding under the \textit{Beyond}Fingerprinting Sandia Grand Challenge Laboratory Directed Research and Development (GC LDRD) program. 

Sandia National Laboratories is a multi-mission laboratory managed and operated by National Technology \& Engineering Solutions of Sandia, LLC (NTESS), a wholly owned subsidiary of Honeywell International Inc., for the U.S. Department of Energy’s National Nuclear Security Administration (DOE/NNSA) under contract DE-NA0003525. This written work is authored by an employee of NTESS. The employee, not NTESS, owns the right, title and interest in and to the written work and is responsible for its contents. Any subjective views or opinions that might be expressed in the written work do not necessarily represent the views of the U.S. Government. The publisher acknowledges that the U.S. Government retains a non-exclusive, paid-up, irrevocable, world-wide license to publish or reproduce the published form of this written work or allow others to do so, for U.S. Government purposes. The DOE will provide public access to results of federally sponsored research in accordance with the DOE Public Access Plan.

\end{acknowledgments}

\section*{Author contributions}
\par{
The design of the objective function, implementation of the Bayesian optimization algorithm, and performance analysis were conducted by AS.
MK and JC conducted the sputter deposition of Mo thin films and measured stress and sheet resistance values. 
HNN and DPA proposed the research idea and supervised the work.
All the authors verified the data and results and contributed to writing and reviewing the manuscript.
}

\section*{Competing interests}
\par{
The authors declare no competing interests.
}
\section*{Materials \& Correspondence}

Requests and correspondence should be addressed to Ankit Shrivastava.

\section*{Data and Code Availability}

The Bayesian optimization code for this work is available at \url{https://github.com/ashriva16/bayesian-optimization-sputter-deposition}.


\appendix
\section{Data from Bayesopt guided experiments}\label{sec:appendix}

This section provides comprehensive information about the BayesOpt guided experiments. 
The values of the proposed configurations $\bar{\bx}$, and optimal configurations $\bx^*$ during Bayesian optimization, discussed \cref{fig:config_trends}, are listed in \cref{tab:trends_deposition}. 
The proposed configurations were rounded off for the experiments, as feasible power values were restricted to multiples of $10$\,W, and feasible pressure values were constrained to increments of $0.5$\,mTorr.
The corresponding feasible configurations used during the experiments are also listed in the table.

At each iteration, several films were deposited using the respective utilized configuration, and their residual stress and sheet resistance were measured, as detailed in Tables \ref{tab:bayesopt_stress} and \ref{tab:bayesopt_resistance}, respectively. 
Note that more than three depositions were performed for iterations $3$ and $4$. 
During the series of experiments, chamber cleaning became necessary due to the accumulation of previously sputtered materials inside the chamber. 
Over time, this buildup can peel off from chamber components and contaminate samples or cause arcing during a deposition. 
The cleaning process can also introduce moisture and other vapors into the chamber, which is known to affect the film properties of depositions immediately following the cleaning\cite{mattox1998preparation}. 
This effect likely explains a larger variation in stress values of iterations $3$ and $4$, which include films deposited before and after chamber cleaning.

\begin{table}[htp]
    \centering
\captionsetup{justification=raggedright,singlelinecheck=false}
    \caption{
    Tabulated data showing trends in configurations observed during the BayesOpt guided search at each iteration.
    The feasible configurations utilized for the experiments are shown in the fourth column.
    }
    \begin{tabular}{C{4pc}C{4.5pc}C{5pc}C{4.5pc}}
        \hline \hline
        {Iteration}  & {Optimal} & {Proposed} & {Utilized} \\
        $\mathbf{i}$ & $\bx^*$  & $\bar{\bx}$ & $\bx$ \\
        \hline \hline
        1 & $(2.0, 750)$ & $(2.0, 648.28)$ & $(2.0, 650)$ \\

        2 & $(2.0, 750)$ & $(2.0, 433.32)$ & $(2.0, 430)$ \\

        3 & $(2.0, 750)$ & $(2.0, 558.49)$ & $(2.0, 560)$ \\

        4 & $(2.0, 650)$ & $(2.0, 667.06)$ & $(2.0, 670)$ \\

        5 & $(2.0, 650)$ & $(2.0, 615.34)$ & $(2.0, 620)$ \\

        6 & $(2.0, 620)$ & $(2.0, 322.11)$ & $(2.0, 320)$ \\

        7 & $(2.0, 320)$ & $(4.35,  323.98)$ & $(4.5, 320)$ \\

        8 & $(2.0, 620)$ & $(18.46, 302.29)$ & $(18.5, 300)$ \\

        9 & $(2.0, 620)$ & $(6.08, 644.67)$ & $(6.0, 650)$ \\

        10 & $(2.0, 620)$ & $(2.0, 236.79)$ & $(2.0, 240)$ \\

        11 & $(2.0, 620)$ & $(12.24, 305.94)$ & Stopped\\
        \hline \hline
    \end{tabular}
    \label{tab:trends_deposition}
\end{table}

\begin{table}[htp]

    \centering
    \captionsetup{justification=raggedright,singlelinecheck=false}
    \caption{Residual stress measurements, along with their computed mean and standard deviation at each iteration of the BayesOpt guided experiments.
    }
    \begin{tabular}{c|c|C{3.5cm}|C{1.2cm}|C{1.2cm}}
        \hline \hline
        $\mathbf{i}$ & \textbf{Utilized} & \textbf{Measurements in MPa} & \textbf{mean} & \textbf{stdev} \\ 
        \hline \hline
        1 & $(2.0, 650)$ & $-123.5, -134.8, -191$ & $-149.8$ & $36.2$  \\
        \hline
        2 & $(2.0, 430)$ & $330.0, 353.5, 349.8$ & $344.4$ & $12.6$\\
        \hline
        3 & $(2.0, 560)$ & $97.9, 40.1, 105.5,$ $28.3, -15.3, 3.2$ & $43.3$ & $49.2$\\
        \hline
        4 & $(2.0, 670)$ & $111.2, -239.6, -171.1,$ $-354, -411.1, -73.9,$ $-276.6, -309.3$ & $-243.4$ & $117.9$\\
        \hline
        5 & $(2.0, 620)$ & $-180.9, -218.0, -215.5$ & $-204.8$ & $20.7$\\
        \hline
        6 & $(2.0, 320)$ & $148.1, 218.6, 164.0$ & $176.9$ & $37.0$\\
        \hline
        7 & $(4.5, 320)$ & $1648.4, 1660.3, 1626.8$ & $1645.2$ & $17.0$\\
        \hline
        8 & $(18.5, 300)$ & $39.7, 38.4, 36.4$ & $38.2$ & $1.7$\\
        \hline
        9 & $(6.0, 650)$ & $1534.3, 1549.20, 1528.03$ & $1537.2$ & $10.9$\\
        \hline
        10 & $(2.0, 240)$ & $766.4, 756.8, 722.1$ & $748.4$ & $23.3$\\        \hline \hline
    \end{tabular}
    \label{tab:bayesopt_stress}
\end{table}

\begin{table}[htp]
    \centering
    \captionsetup{justification=raggedright,singlelinecheck=false}
    \caption{Sheet resistance measurements, along with their computed mean and standard deviation at each iteration of the BayesOpt guided experiments.
    }
    \begin{tabular}{c|c|C{3.5cm}|C{1.2cm}|C{1.2cm}}
        \hline \hline
        $\mathbf{i}$ & \textbf{Utilized} & \textbf{Measurements in $\runit$} & \textbf{mean} & \textbf{stdev} \\ 
        \hline \hline
        1 & $(2.0, 650)$ &  $0.87, 0.86, 0.85$ & $0.86$ & $0.01$  \\
        \hline
        2 & $(2.0, 430)$ & $0.74, 0.74, 0.73$ & $0.74$ & $0.01$ \\
        \hline
        3 & $(2.0, 560)$ &  $0.64, 0.64, 0.63,$ $0.84, 0.78, 0.81$ & $0.72$ & $0.10$ \\
        \hline
        4 & $(2.0, 670)$ & $0.72, 0.66, 0.70,$ $0.65, 0.65, 0.66,$ $0.74, 0.72$ & $0.69$ & $0.04$ \\
        \hline
        5 & $(2.0, 620)$ & $0.68, 0.68, 0.77$ & $0.71$ & $0.05$ \\
        \hline
        6 & $(2.0, 320)$ & $1.44, 1.44, 1.42$ & $1.43$ & $0.01$ \\
        \hline
        7 & $(4.5, 320)$ & $1.68, 1.68, 1.7$ & $1.69$ & $0.01$\\
        \hline
        8 & $(18.5, 300)$ & $8.04, 7.9, 8.14$ & $8.03$ & $0.12$\\
        \hline
        9 & $(6.0, 650)$ & $1.68, 1.68, 1.70$ & $1.69$ & $0.01$ \\
        \hline
        10 & $(2.0, 240)$ & $0.81, 0.82, 0.85$ & $0.83$ & $0.02$ \\
        \hline \hline
    \end{tabular}
    \label{tab:bayesopt_resistance}
\end{table}

\section{Data from experiments for optimum verification}\label{sec:appendix_additional}

This appendix provides comprehensive information about the additional experiments conducted for verification of the optimum solution obtained by Bayesian optimization. 
The experiments were conducted with configurations near the optimal configuration to observe the sensitivity of residual stress dependence on pressure to process variations. 
For each configuration, three films were deposited, and the residual stress and sheet resistance were measured as shown in Tables~\ref{tab:add_stress} and~\ref{tab:add_resistance}, respectively.

Note that small differences in the stress measurements were observed at $(2.0~\text{mTorr},~620~\text{W})$ as compared to the corresponding measurements taken during the BayesOpt-guided experiments.
Given the significant time gap between experiments, several factors may have influenced the discharge voltage of the sputter target. 
These include small target impedance changes when the target is installed and re-installed, the number of kWh the target has been used, and small pressure differences that are outside what the capacitance manometer on the system can measure.  
Nevertheless, thin-film stress and resistivity exhibit similar trends with power and pressure despite minor deviations in absolute values. Additionally, all design criteria continue to be satisfied at the optimum configuration proposed by BayesOpt.  

\begin{table}[htp]

    \centering
    \captionsetup{justification=raggedright,singlelinecheck=false}
    \caption{
    Measurements of residual stress, along with their computed mean and standard deviation during the additional experiments.   
    }
    \begin{tabular}{c|C{3.7cm}|C{1.2cm}|C{1.2cm}}
        \hline \hline
        \textbf{Utilized} & \textbf{Measurements in MPa} & \textbf{mean} & \textbf{stdev} \\ 
        \hline \hline
        $(2.0, 590)$ & $-64.50, -84.73, -107.52$ & $-85.58$ & $21.52$ \\
        \hline
        $(2.5, 590)$ & $382.83, 409.49, 414.19$ & $402.17$ & $16.91$ \\
        \hline
        $(2.0, 620)$ & $-52.34, -35.65, -19.57$ & $-43.99$	& $16.39$ \\
        \hline
        $(2.5, 620)$ & $140.26, 175.44, 151.19$ & $155.63$ & $18.01$ \\
        \hline
        $(3.0, 620)$ & $679.36, 691.20, 641.25$ & $670.60$ & $26.10$ \\
        \hline
        $(2.0, 650)$ & $-303.79, -254.90, -211.19$ & $-256.63$ & $46.32$\\
        \hline
        $(2.5, 650)$ & $292.57, 314.78, 295.83$ & $301.06$ & $11.99$ \\
        \hline \hline
    \end{tabular}
    \label{tab:add_stress}
\end{table}

\begin{table}[htp]
    \centering
    \captionsetup{justification=raggedright,singlelinecheck=false}
    \caption{
    Measurements of sheet resistance, along with their computed mean and standard deviation during the additional experiments.  
    }
   \begin{tabular}{c|C{3.5cm}|C{1.2cm}|C{1.2cm}}
        \hline \hline
        \textbf{Utilized} & \textbf{Measurements in $\runit$} & \textbf{mean} & \textbf{stdev} \\ 
        \hline \hline
        $(2.0, 590)$ & $0.63, 0.61, 0.62$ & $0.62$& $0.01$ \\
        \hline
        $(2.5, 590)$ & $0.74, 0.71, 0.72$ & $0.72$& $0.01$ \\
        \hline
        $(2.0, 620)$ & $0.78, 0.77, 0.76$ & $0.77$& $0.01$ \\
        \hline
        $(2.5, 620)$ & $0.62, 0.61, 0.64$ & $0.62$& $0.01$ \\
        \hline
        $(3.0, 620)$ & $0.66, 0.65, 0.67$ & $0.66$& $0.01$ \\
        \hline
        $(2.0, 650)$ & $0.58, 0.57, 0.57$ & $0.57$& $0.01$ \\
        \hline
        $(2.5, 650)$ & $0.61, 0.59, 0.60$ & $0.60$& $0.01$ \\
        \hline \hline
    \end{tabular}
    \label{tab:add_resistance}
\end{table}

\section*{References}
\bibliography{bibliography}

\end{document}